%% file: main.tex
\documentclass[a4paper,fleqn]{cas-dc}

\usepackage[authoryear]{natbib}

\input{preamble.tex}

\begin{document}
\let\WriteBookmarks\relax
\def\floatpagepagefraction{1}
\def\textpagefraction{.001}
\shorttitle{Computationally efficient Gauss-Newton RL for MPC}
\shortauthors{D. Brandner, S. Gros, and S. Lucia}

\title [mode = title]{Computationally efficient Gauss-Newton reinforcement learning for model predictive control}

\nonumnote{This work was funded by the Deutsche Forschungsgemeinschaft (DFG, German Research Foundation) – 466380688 – within the Priority Program “SPP 2331: Machine Learning in Chemical Engineering”.}

\author[1]{Dean Brandner}
\cormark[1]

\ead{dean.brandner@tu-dortmund.de}

\credit{Writing - Original Draft, Writing - Review \& Editing, Visualization, Software, Conceptualization, Methodology, Investigation}

\affiliation[1]{organization={Chair of Process Automation Systems, TU Dortmund University},
                addressline={Emil-Figge-Str. 70}, 
                city={Dortmund},
                postcode={44227}, 
                country={Germany}}

\author[2]{Sebastien Gros}
\ead{sebastien.gros@ntnu.no}
\affiliation[2]{organization={Department of Engineering Cybernetics, Norwegian University of Science and Technology (NTNU)},
                addressline={H\o gskoleringen 1}, 
                city={Trondheim},
                postcode={7034}, 
                country={Norway}}

\credit{Writing - Review \& Editing, Methodology, Supervision}

\author[1]{Sergio Lucia}
\ead{sergio.lucia@tu-dortmund.de}

\credit{Writing - Review \& Editing, Conceptualization, Methodology, Supervision, Funding acquisition}

\let\printorcid\relax

\begin{abstract}
Model predictive control (MPC) is widely used in process control due to its interpretability and ability to handle constraints.
As a parametric policy in reinforcement learning (RL), MPC offers strong initial performance and low data requirements compared to black-box policies like neural networks.
However, most RL methods rely on first-order updates, which scale well to large parameter spaces but converge at most linearly, making them inefficient when each policy update requires solving an optimal control problem, as is the case with MPC.
While MPC policies are typically \reviewed{low} parameterized and thus amenable to second-order approaches, existing second-order methods demand second-order policy derivatives, which can be computationally intractable.

This work introduces a Gauss-Newton approximation of the deterministic policy Hessian that eliminates the need for second-order policy derivatives, enabling superlinear convergence with minimal computational overhead.
To further improve robustness, we propose a momentum-based Hessian averaging scheme for stable training under noisy estimates \reviewed{coupled with an adaptive trust-region.}
We demonstrate the effectiveness of the approach on a nonlinear continuously stirred tank reactor (CSTR), showing faster convergence and improved data efficiency over state-of-the-art first-order methods \reviewed{and deep RL approaches}.
\end{abstract}

%\begin{highlights}
%\item Gauss-Newton approximation avoids second-order policy derivative computation
%\item Ensures superlinear convergence with low computational overhead
%\item Momentum averaging improves Hessian stability under noisy estimates
%\item Data-efficient learning with reduced wall-clock time on nonlinear CSTR case study
%\end{highlights}


\maketitle

\section{Introduction}
Model predictive control~(MPC) is a popular approach for the control of highly dynamic, nonlinear multivariable systems subject to safety and quality constraints.
In MPC, a control plant system model is used to predict and optimize the future behavior of the system states so that an objective function is minimized while \secondrevision{enforce} process constraints.
The first control action of the optimal solution is then deployed to the real plant, the next state is measured or estimated, and the process is repeated.
Since MPC delivers a sequence of optimal control inputs and a sequence of upcoming states, the control policy is highly interpretable.
As an established method, many theoretical results regarding stability and feasibility exist~\citep{rawlingsModelPredictiveControl2017}.
Besides rigorous model-based optimal control ap\-proa\-ches like MPC, there are emerging data-driven approaches such as reinforcement learning~(RL) that show suitability and success in the process industry~\citep{yooReinforcementLearningBatch2021, gopaluniModernMachineLearning2020, leeIterativeLearningControl2007}.
Model-free RL as a subfield RL focuses on learning an optimal policy through trial-and-error interaction between an agent (controller) and its environment (plant), without relying on a model of the environment.
In practice, this optimal policy is often approximated directly using a rich parametric function approximator such as neural networks~(NN).
Contrary to MPC, this has the benefit that no differentiable system model is necessary to perform online optimization, which is especially interesting if models are hard to build or impossible to differentiate.
However, in general model-free RL approaches need large amounts of data, which can be prohibitive for application to industrial scale processes~\citep{rechtTourReinforcementLearning2019}.

We focus in this work on two main reasons for the large data requirements of RL: the difficulty of initialization and the use of first-order optimization approaches.
If expert knowledge is available beforehand, the NN policy of the RL agent can be pretrained to deliver a reasonable initial performance.
However, if expert knowledge is not available, the NN policy is often initialized with random values such that many updates are necessary to \secondrevision{reach an acceptable performance.}
Most popular model-free RL algorithms only consider update rules that consist of first-order derivative information such as~\citet{fujimotoAddressingFunctionApproximation2018} and~\citet{haarnojaSoftActorCriticOffPolicy2018}.
In either case, the maximum convergence rate of first-order approaches is limited to be linear~\citep{bottouOptimizationMethodsLargeScale2018}.
Methods that consider (approximate) second-order derivative information such as Newton, quasi-Newton, or inexact Newton methods can provide up to quadratic convergence.
However, all those approaches must store at least a Hessian matrix approximation in memory and a linear system of equations must be solved for each update step at least approximately.
This can be prohibitive especially when rich parameterized policy approximators such as NNs are used.
Works such as~\citet{korbitExactGaussNewtonOptimization2024}, \citet{liuSophiaScalableStochastic2024} and \citet{yaoADAHESSIANAdaptiveSecond2021} address the memory issue by approximating the Hessian via scalars or diagonal representations, whereas works such as~\citet{chauhanStochasticTrustRegion2020} and \citet{schraudolphStochasticQuasinewtonMethod2007} only solve the linear system of equations inexactly.

\begin{figure*}
    \centering
    \includegraphics[width =\textwidth]{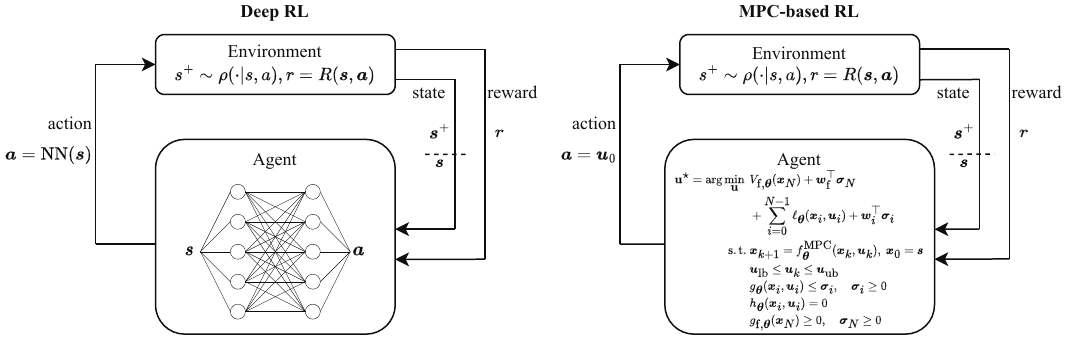}
    \caption{Comparison between deep RL (left) and MPC-based RL (right).}
    \label{fig:Comparison_DeepRL_vs_MPC-based_RL}
\end{figure*}

Recent research identifies the challenges described above and proposes to use MPC instead of NNs as a policy for the RL agent~\citep{grosDataDrivenEconomicNMPC2020,bankerGradientBasedFrameworkBilevel2025, brandnerReinforcedModelPredictive2024b, mayfrankEndtoendReinforcementLearning2024}.
Figure~\ref{fig:Comparison_DeepRL_vs_MPC-based_RL} illustrates how NN policies can be replaced by MPC.
\citet{reiterSynthesisModelPredictive2025} provide a comprehensive review for MPC used as policy in RL.
The fundamental idea is to introduce tunable parameters to the MPC's optimal control problem and to adjust them in the same way as the parameters in NN policies.
Contrary to NN policies, MPC provides a reasonable initial policy \secondrevision{by solving a model-based optimal control problem that directly converts system knowledge into control actions.}
\reviewed{It also provides a universal approximator to the solution of Markov decision processes~(MDP), which states the theoretical framework for RL~\citep{grosDataDrivenEconomicNMPC2020}.}
Further, MPC typically has fewer tuning parameters than NN policies.
Due to the \reviewed{low} parameterization of MPC policies, Hessian computation becomes tractable, allowing more sophisticated updates with superlinear convergence rates, which are of significant interest for MPC policies because each RL transition requires the solution of an optimization problem, which is computationally expensive.

Different update rules were proposed in the past to improve data efficiency.
\citet{Kakade_NaturalPolicyGradient_2001} introduces the natural policy gradient, which scales the policy gradient by the inverse of the Fisher information matrix.
Although being limited to linear convergence, it often shows overall practical improvement.
Further improvement is obtained via quasi-Newton approaches with a superlinear convergence rate as proposed in \citet{furmstonApproximateNewtonMethods2016}, \citet{jhaQuasiNewtonTrustRegion2020}, and \citet{kordabadQuasiNewtonIterationDeterministic2022}.
With a special focus on MPC policies, \citet{brandnerReinforcedModelPredictive2024b} propose a tailored algorithm that reduces the number of updates significantly.
However, the second-order policy derivative \reviewed{of the MPC policy} with respect to its parameters needs to be computed.
\citet{brandnerReinforcedModelPredictive2024b} show that the second-order derivative of MPC policies can be obtained by the solution of a linear system of equations.
Unfortunately, the right-hand side of the linear system involves third-order derivatives of the MPC's Lagrangian with respect to all primal-dual variables, which can be computationally demanding and memory intensive, especially for high dimensional and highly nonlinear system models.
Lastly, all considered update rules, regardless if first or second-order, require to compute expected values and further assume perfect knowledge of the action-value function, which evaluates how good a certain action in a given state is, when following the present policy.
In practice however, these expected values and the action-value function can only be approximated.
Both factors contribute to the fact that the gradient and the Hessian cannot be computed perfectly, which can lead to unstable training.

In this work, we address both the high memory requirements \reviewed{as well as the} high computational demand when using MPC policies in second-order RL due to the second-order policy derivatives, and the training instabilities due to poor gradient and Hessian estimation.
Our main contributions are twofold:
\begin{enumerate}[1)]
    \item
    We propose a Gauss-Newton approximation of the deterministic policy Hessian that does not require second-order policy derivatives, leading to significant computational savings. We rigorously prove that our approach can still lead to superlinear convergence \reviewed{due to faster contraction} despite the simpler required computations.
    Our proposed approximation alleviates the high data requirements compared to the established first-order policy updates in RL.
    \item
    We derive a momentum-based update rule for Hessian estimation that returns an unbiased exponentially averaged Hessian estimate.
    This estimate is robust with respect to noise in local Hessian samples and therefore enables stable and fast training as it reduces the noise in the eigenvalues of the Hessian.
    A reliable Hessian estimate is very relevant for MPC policies in RL because the influence of the policy parameters on the objective can typically vary by several orders of magnitude.
    \reviewed{We embed this unbiased exponentially averaged Hessian estimate of the Gauss-Newton approximation within an adaptive trust-region constrained optimization problem to compute robust parameter updates.}
\end{enumerate}

The work is structured as follows.
We introduce the relevant preliminaries on Markov decision processes, iterative policy optimization via RL, and MPC as policy approximator in RL in Section~\ref{seq:preliminaries}.
We present the proposed Gauss-Newton Hessian approximation, the momentum-based update rule, and the adaptive trust-region constrained update computation in Section~\ref{sec:Contribution}.
We illustrate the superlinear convergence capability of the Gauss-Newton approximation on an analytical case study in Section~\ref{sec:Analytical_Case_Study} and show the practical improvement of the \reviewed{proposed approach} at the example of a nonlinear continuously stirred tank reactor~(CSTR) in Section~\ref{sec:CSTR}.
Section~\ref{sec:Conclusion} concludes the work and provides perspectives for future work.

\section{Preliminaries} \label{seq:preliminaries}

\subsection{Markov decision processes}
RL can solve problems that can be formalized as a MDP.
MDPs are a general mathematical modeling framework for sequential decision making~\citep{suttonReinforcementLearningIntroduction2018}.
In an MDP, an agent (controller) interacts with an environment (plant) as depicted in Figure~\ref{fig:Comparison_DeepRL_vs_MPC-based_RL}.
The agent computes a certain action~$\va\in\calA\subseteq\sR^{n_{\va}}$ according to the agent's decision making policy~$\pi$, given the current environmental state~$\vs\in\calS\subseteq\sR^{n_{\vs}}$.
In this work, we focus on deterministic polices~$\pi:\calS \rightarrow \calA$ that directly map the current state~$\vs$ to its proposed action~$\va$.
This action~$\va$ is handed to the environment, which transitions from the current state~$\vs$ to a subsequent state~$\vs^+\in \calS$ according to the MDP's conditional transition probability~$\rho: \calS \times \calS \times \calA \rightarrow [0 , \infty)$.
The subsequent state is sampled from~$\rho$ according to
\begin{align}
    \vs^+ \sim \rho(\cdot \vert \vs, \va). \label{eq:MDP_transition_prob}
\end{align}
Once that the environment transitioned from state~$\vs$ to $\vs^+$, the new environmental state is handed to the agent together with a reward~$r\in\sR$.
This reward~$r$ is computed according the environment's reward function~$R:\calS\times\calA \rightarrow \sR$, which depends on the current state~$\vs$ and the taken action~$\va$, and serves as an immediate measure of how good the current state-action pair \reviewed{$\left(\vs, \va \right)$} is, while neglecting future consequences due to upcoming subsequent states.

The general goal under which an MDP is solved, is to find a policy that maximizes not only the immediate reward but the expected future cumulative reward.
To measure the expected future cumulative reward in a given state, the concept of value functions are used.
The state-value function~$V^{\pi}:\calS \rightarrow \sR$ measures how good policy~$\pi$ performs when being in state~$\vs$.
It is defined as the expected sum of rewards when being in state~$\vs$ and taking the action proposed by the policy~$\va = \pi(\vs)$
\begin{align}
    &\begin{aligned}
        V^{\pi}(\vs) =& R(\vs, \pi(\vs)) + \gamma \sE_{\vs^+} \left[R(\vs^+, \pi(\vs^+))\right] \notag \\
        &+ \gamma^2 \sE_{\vs^+}\left[\sE_{\vs^{++}}\left[R(\vs^{++}, \pi(\vs^{++})) \right]\right] + \ldots \notag%
    \end{aligned}
    \\&= \sE \left[ \left.\sum_{i = 0}^\infty \gamma^i R(\vs_i, \pi(\vs_i)) \right \vert \vs_0 = \vs, \vs_{i+1} \sim \rho(\cdot\vert\vs_i,\pi(\vs_i))\right]
\end{align}
with $0 < \gamma \leq 1$ being a discount factor that allows to balance immediate and future reward.
All expected values are taken over the conditional transition probability~$\rho(\cdot \vert \vs, \pi(\vs))$ under closed-loop operation.
The state-value function can also written recursively as
\begin{align}
    V^{\pi}(\vs) = R(\vs, \pi(\vs)) + \gamma \sE_{\vs^+ \sim \rho(\cdot\vert\vs, \pi(\vs))}\left[ V^{\pi}(\vs^+)\right]. \label{eq:state_value_function_recursively}
\end{align}
The action-value function~$Q^\pi:\calS \times \calA \rightarrow \sR$ generalizes the state-value function to taken actions that are different from actions proposed by the policy.
Analogous to the state-value function, the action-value function is recursively defined as
\begin{align}
    Q^\pi(\vs,\va) = R(\vs, \va) + \gamma \sE_{\vs^+ \sim \rho(\cdot \vert \vs, \va)} \left[ V^\pi (\vs^+)\right].
\end{align}
In general, the policy should be optimal not only for one single state~$\vs$ but for all possible states.
To generalize the local state-dependent policy performance measured by~$V^\pi(\vs)$ to all possible states that can occur under the MDP, the expected cumulative reward~$J$ is introduced.
It is defined as the expected value of the state-values for all initial states~$\vs_0 \in \calS_0 \subseteq\calS$ that can occur when being sampled from the distribution~$\rho_0:\calS_0 \rightarrow [0, \infty)$
\begin{align}
    J(\pi) = \sE_{\vs_0\sim \rho_0} \left[V^{\pi}(\vs_0)\right]. \label{eq:Definition_CLC}
\end{align}

A policy~$\pi^\star$ is considered optimal (denoted with $^\star$) when it maximizes the expected cumulative reward
\begin{align}
    \pi^\star = \arg \max_{\pi} J(\pi). \label{eq:functional_policy_optimization}
\end{align}
The following inequalities can be derived directly from the optimality of the policy~$\pi^\star$
\begin{align}
    V^{\pi^\star}(\vs) &\geq V^{\pi}(\vs), \quad \forall \vs \in \calS.\\
    Q^{\pi^\star}(\vs, \pi^\star(\vs)) &\geq Q^{\pi^\star}(\vs, \va), \quad \forall \vs, \va \in \calS \times \calA. \label{eq:optimal_action_value_function}
\end{align}

\subsection{Iterative policy optimization in reinforcement learning} \label{seq:PolicyOptimization}
To find the policy~$\pi^\star$ that solves~\eqref{eq:functional_policy_optimization}, an infinite dimensional optimization problem over the space of all admissible policy functions must be solved.
This is intractable in most cases.
Therefore, the policy is often parameterized by a function approximator~$\pi_{\vtheta}(\vs)$, where~$\vtheta\in\sR^{n_{\vtheta}}$ are tunable policy parameters~\citep{suttonReinforcementLearningIntroduction2018}.
The expected cumulative reward transforms from a function of the policy~$J(\pi)$ into a function of the policy parameters~$J(\pi_{\vtheta}) = J(\vtheta)$.
The infinite dimensional optimization problem~\eqref{eq:functional_policy_optimization} simplifies to the finite dimensional optimization problem
\begin{align}
    \vtheta^\star = \arg \max_{\vtheta} J(\vtheta). \label{eq:parametric_policy_optimization}
\end{align}

In its general form, optimization problem~\eqref{eq:parametric_policy_optimization} can be categorized as an unconstrained nonlinear program with stochastic objective since $J$ is defined in terms of expected values as shown in~\eqref{eq:Definition_CLC}.
Iterative optimization algorithms are standard tools to solve such optimization problems numerically~\citep{nocedalNumericalOptimization2006}.
The basic idea is to start at an initial guess~$\vtheta_0$ of the optimal solution and improve the guess iteratively.
The general iterative update process at the $k$-th iteration step can be described as
\begin{align}
    \vtheta_{k+1} = \vtheta_{k} + \vp_k
\end{align}
with $\vp_k\in\sR^{n_{\vtheta}}$ denoting the applied parameter update.
The way that the parameter update~$\vp_k$ is computed depends on the chosen update rule, which severely influences the performance of the iterative optimization.
Different update rules will be presented in the following subsections.
The iteration process is repeated until the iterates converge to a point that satisfies the first-order optimality condition for unconstrained optimization sufficiently well
\begin{align}
    \left\|\nabla_{\vtheta} J(\vtheta)\right\| \leq \epsilon \label{eq:Optimality_Unconstrained}
\end{align}
where $\|\cdot\|$ denotes a vector-norm and~$\epsilon \geq 0$ is a constant.

\subsubsection{First-order policy optimization} \label{sec:first_order_optimization}
An optimization algorithm is considered to be of first-order if it only uses first-order derivative information of the optimization problem during the optimization process.
In case of~\eqref{eq:parametric_policy_optimization}, this refers to the gradient of the expected cumulative reward with respect to the policy parameters $\nabla_{\vtheta} J(\vtheta) \in \sR^{n_{\vtheta}}$.
In order to preserve a concise and clutter-free notation, we introduce the gradient of the objective function, evaluated at the current iterate~$\vtheta_k$ as $\vg_k = \nabla_{\vtheta} J(\vtheta)\vert_{\vtheta = \vtheta_k}$.
\citet{silverDeterministicPolicyGradient2014} derive the deterministic policy gradient as 
\begin{align}
    \nabla_{\vtheta} J(\vtheta) = \sE_{\vs} \left[ \nabla_{\vtheta} \pi_{\vtheta}^\top (\vs) \nabla_{\va} Q^{\pi_{\vtheta}}(\vs, \va)\vert_{\va = \pi_{\vtheta}(\vs)} \right] \label{eq:DPG}
\end{align}
where $\nabla_{\vtheta}\pi_{\vtheta}(\vs)\in\sR^{n_{\va} \times n_{\vtheta}}$ denotes the Jacobian of the policy with respect to its parameters and $\nabla_{\va}Q^{\pi_{\vtheta}}(\vs, \va)\in \sR^{n_{\va}}$ denotes the gradient of the policy's action-value function with respect to the taken actions, which is evaluated at the policy's proposed action~$\va = \pi_{\vtheta}(\vs)$.
\reviewed{The expectation~$\sE_{\vs}[\cdot]$ denotes the expectation over the discounted sum of the Markov chain that occurs under closed-loop operation with policy~$\pi_{\vtheta}$.}

As the simplest form of a first-order method for stochastic objectives, we introduce the stochastic gradient ascent algorithm~\citep{robbinsStochasticApproximationMethod1951}.
The stochastic gradient ascent algorithm stems from the deterministic gradient ascent algorithm, which updates the current iterate~$\vtheta_k$ into the direction of steepest ascent, which is the gradient of the objective function at the current iterate~$\vg_k$.
Since the deterministic policy gradient~\eqref{eq:DPG} is computed using an expected value, it can only be estimated empirically using a finite amount of samples \reviewed{from $N_\mathrm{IC} \in \sN$ initial conditions, each having episodes a length of $N_{\mathrm{ep},i} \in \sN$} such that
\reviewed{%
\begin{align}
    \tilde{\vg}_k = \frac{1}{N_\mathrm{IC}} \sum_{i = 1}^{N_\mathrm{IC}} \sum_{j=0}^{N_{\mathrm{ep}, i}} \gamma^j \nabla_{\vtheta} \pi_{\vtheta}^\top (\vs_{i,j}) \nabla_{\va} Q^{\pi_{\vtheta}}(\vs_{i,j}, \va)\vert_{\va = \pi_{\vtheta}(\vs_{i,j})} \label{eq:finite_DPG}%
\end{align}%
}
with $\tilde{\vg}_k \approx \vg_k$ leading to noisy estimates of the true deterministic policy gradient.
The parameter update~$\vp_k$ for the stochastic gradient ascent algorithm is computed as
\begin{align}
    \vp_k = \alpha \tilde{\vg}_k \label{eq:SGD}
\end{align}
where $\alpha > 0$ is a scalar step size, also often called the learning rate.

Many works aim to improve the stability and the convergence of stochastic gradient ascent by applying an exponential moving average to \reviewed{the sampled} gradient estimates~\citep{kingmaAdamMethodStochastic2015,Duchi_2011_AdaGrad, rumelhartLearningRepresentationsBackpropagating1986}.
These methods are called momentum-based methods.
In its core, momentum-based methods update an exponential moving average~$\vm_k \in \sR^{n_{\vtheta}}$ of past gradients with the newly obtained gradient estimate~$\tilde{\vg}_k$ according to
\begin{align}
    \vm_k = \reviewed{\beta_1} \vm_{k-1} + (1 - \reviewed{\beta_1}) \tilde{\vg}_k \label{eq:gradient_estimation_w_momentum}
\end{align}
where $0 \leq \reviewed{\beta_1} \leq 1$ is an exponential decay factor. %
The larger~$\reviewed{\beta_1}$, the slower the moving average~$\vm_k$ adapts to the current gradient estimate. %
Typically, the moving average is initialized as a zero vector~$\vm_0 = 0$.
This causes initialization bias of the exponential moving average towards zero, which results in slow improvement in early stages of the optimization.
\citet{kingmaAdamMethodStochastic2015} correct the initialization bias by
\begin{align}
    \hat{\vm}_k = \frac{\vm_k}{(1 - \reviewed{\beta_1^k})} \label{eq:unbiasing_of_gradient_estimation_w_momentum}
\end{align}
with $\hat{\vm}_k \in \sR^{n_{\vtheta}}$ being the bias-corrected exponential moving average.
The update rule of a the bias-corrected momentum-based stochastic gradient ascent method is similar to~\eqref{eq:SGD} but replaces the gradient with the bias-corrected average
\begin{align}
    \vp_k = \alpha \hat{\vm}_k. \label{eq:SGD_w_momentum}
\end{align}

\reviewed{
The entries of the bias-corrected exponential moving gradient average~$\hat{\vm}_k$ can still range over several orders of magnitude, which can cause rapid learning in directions with large components and small updates into directions with small entries.
The state-of-the-art first-order optimizer Adam~\citep{kingmaAdamMethodStochastic2015} tackles this challenge by rescaling all components by an exponential moving average of the magnitude of each individual component.
For this, the second exponential moving average $\vv_k\in\sR^{n_{\vtheta}}$ is introduced.
It captures the square of each gradient entry as 
\begin{align}
    \vv_k = \beta_2 \vv_{k-1} + (1 - \beta_2) \tilde{\vg}_k \odot \tilde{\vg}_k \label{eq:Adam_second_order_moment}
\end{align}
with $0 \leq \beta_2 \leq 1$ and $\odot$ denoting the Hadamard (element-wise) product.
The introduced initialization bias is corrected similar to~\eqref{eq:unbiasing_of_gradient_estimation_w_momentum}.
Finally, the improved update rule of the Adam optimizer is defined as
\begin{align}
    \vp_{k} = \alpha \frac{\hat{\vm}_k}{\sqrt{\hat{\vv}_k} + \epsilon} \label{eq:Adam_update}
\end{align}
where $\epsilon > 0$ is a constant, introduced for numerical stability, and the division operation is meant to be element-wise.
}

\subsubsection{Second-order policy optimization}
Second-order methods improve upon first-order methods by also considering second-order derivative information.
The commonly used Newton method derives the following update rule
\begin{align}
    \vp_k = - \mH_k^{-1} \vg_k \label{eq:exact_newton}
\end{align}
where $\mH_k$ denotes the general Hessian matrix of the objective function with respect to its decision variables at the $k$-th iterate.
It can be shown that this update rule delivers a quadratic convergence rate~\citep{nocedalNumericalOptimization2006}.
In case of~\eqref{eq:parametric_policy_optimization}, $\mH_k$ refers to the Hessian of the expected cumulative reward with respect to the policy parameters~$\nabla_{\vtheta}^2J(\vtheta) \in \sR^{n_{\vtheta} \times n_{\vtheta}}$ evaluated at the $k$-th iterate~$\mH_k = \nabla_{\vtheta}^2 J(\vtheta)\vert_{\vtheta = \vtheta_k}$.
However, the computation of the exact deterministic policy Hessian is computationally expensive and therefore prohibitive for most applications.

Quasi-Newton methods are a computationally tractable alternative to exact Newton methods not only for RL but for optimization in general.
The main idea of quasi-Newton methods is to approximate the exact Hessian with a computationally tractable alternative~$\mB_k \approx \mH_k$ at the expense of a reduced superlinear convergence rate~\citep{nocedalNumericalOptimization2006}.
In alignment to Newton methods, the exact Hessian is interchanged with the tractable approximation resulting in the following update rule for quasi-Newton methods
\begin{align}
    \vp_k = - \mB_k^{-1} \vg_k. \label{eq:Quasi-Newton-Rule}
\end{align}
With a focus on policy optimization via RL, \citet{kordabadQuasiNewtonIterationDeterministic2022} propose a computationally cheaper approximation of the deterministic policy Hessian.
The authors show that the following expression approximates the deterministic policy Hessian~$\mM_k =\mB_k \approx \mH_k$ such that still a superlinear convergence rate can be achieved
\begin{align}
    \mM(\vtheta)  = \mM_1(\vtheta) + \mM_2(\vtheta) \label{eq:DPH_approx}
\end{align}
where the matrices $\mM_1(\vtheta), \mM_2(\vtheta)\in \sR^{n_{\vtheta}\times n_{\vtheta}}$ are defined as 
\begin{subequations}
    \begin{align}
    \mM_1(\vtheta) &= \sE_{\vs}\left[ \nabla_{\vtheta}^2 \pi_{\vtheta}(\vs) \otimes \nabla_{\va} Q^{\pi_{\vtheta}}(\vs, \va)\vert_{\va = \pi_{\vtheta}(\vs)}\right]\label{eq:DPH_part_1}\\
    \mM_2(\vtheta) &= \sE_{\vs}\left[\nabla_{\vtheta}\pi_{\vtheta}^\top(\vs) \nabla_{\va}^2 Q^{\pi_{\vtheta}}(\vs, \va)\vert_{\va=\pi_{\vtheta}(\vs)} \nabla_{\vtheta} \pi_{\vtheta}(\vs) \right] \label{eq:DPH_part_2}
    \end{align}
\end{subequations}
with $\nabla_{\vtheta}^2\pi_{\vtheta}(\vs) \in \sR^{n_{\va} \times n_{\vtheta} \times n_{\vtheta}}$ being the second-order Jacobian tensor of the policy with respect to the policy parameters, $\nabla_{\va}^2 Q^{\pi_{\vtheta}}(\vs, \va) \in \sR^{n_{\va} \times n_{\va}}$ denoting the Hessian of the action-value function with respect to the taken action, and the operator~$\otimes:\sR^{n_1 \times n_2 \times n_3} \times \sR^{n_1} \rightarrow \sR^{n_2 \times n_3}$ denoting the tensor-vector product, which we define as
\begin{align}
    \tT \otimes \vv = \sum_{i = 1}^{n_1} \tT_{[i, :, :]} \vv_i
\end{align}
with $\tT_{[i, :,:]}\in \sR^{n_2 \times n_3}$ denoting the $i$-th matrix slice of the general tensor~$\tT \in \sR^{n_1 \times n_2 \times n_3}$ and $\vv_i \in \sR$ denoting the $i$-th element of the general vector $\vv \in \sR^{n_1}$.

In the context of RL, the policy Hessian approximation~\eqref{eq:DPH_approx} is computed using an expected value like the policy gradient.
This expected value must be estimated empirically with a finite amount of samples similar to~\eqref{eq:finite_DPG}, resulting in~$\tilde{\mB}_k \approx \mB_k$.
Subsampled Newton methods form the subcategory of Newton methods that consider optimization with noisy gradient~$\tilde{\vg}_k$ and Hessian estimates~$\tilde{\mB}_k$.
Detailed reviews on subsampled Newton methods are provided in~\citet{bollapragadaExactInexactSubsampled2019} and~\cite{bottouOptimizationMethodsLargeScale2018}.
Analogous to~\eqref{eq:SGD} and~\eqref{eq:Quasi-Newton-Rule}, the update rule for subsampled Newton methods is derived as 
\begin{align}
    \vp_k = - \tilde{\mB}_k^{-1} \tilde{\vg}_k. \label{eq:Subsampled_Newton}
\end{align}
One central idea in subsampled Newton methods is that the quality of the obtained update in~\eqref{eq:Subsampled_Newton} is mainly determined by the accuracy of the gradient estimate~$\tilde{\vg}_k$.
The accuracy of the Hessian estimate only has a minor influence on the proposed update rule as long as it is of sufficient quality.
In many applications the Hessian estimate can be improved by taking large enough sample sizes.
In the context of general large-scale optimization, approaches like AdaHessian~\citep{yaoADAHESSIANAdaptiveSecond2021} or Sophia~\citep{liuSophiaScalableStochastic2024} aim to alleviate instabilities due to insufficient Hessian estimation with the use of momentum on the Hessian estimate.
Both approaches constrain themselves to vectorial Hessian approximations as they are particularly designed for large scale optimization and do not consider a good initialization of the exponentially averaged Hessian estimate.

\reviewed{
Still, even if the sample size is large enough, the Hessian can be singular.
Even if the Hessian is regularized before the Newton update is computed, the update direction can be severely distorted, suggesting large steps into the directions with local linear behavior.
Trust-region approaches constrain the length of the computed update to a certain trust-region radius~$\delta_k > 0$, which can introduce a higher level of stability.
The goal is then to find the optimal step within this trust-region.
In order to compute the trust-region constrained update step~$\vp_{k}$ the following optimization problem must be solved~\citep{nocedalNumericalOptimization2006}, which results from a quadratic approximation of the objective function around the current iterate~$\vtheta_k$
\begin{subequations}
    \begin{align}
        \vp_{k} = \arg \min_{\vp} \quad & \vp^\top \tilde{\vg}_k + \frac{1}{2} \vp^\top \tilde{\mB}_k \vp \\
        \mathrm{s.t.} \quad & \left\|\vp\right\|_2 \leq \delta_k.
    \end{align} 
\end{subequations}
The trust-region radius~$\delta_k$ does not need to be constant and can be adapted depending on how well the objective function and its second-order approximation align.
However, these comparisons are often impractical for stochastic objective functions as they are in RL because the objective estimates are noisy, rendering the comparison unreliable.
Therefore, we will suggest a systematic approach to update the trust-region radius without comparison to the objective function.
}

\subsection{Parameterized model predictive control as policy in reinforcement learning}
RL agents act according to their policy.
Most works in the area of RL focus on approximating the optimal policy using NNs \reviewed{nowadays}.
Despite having a good interpretability, linear functions are often limited in their expressiveness.
NNs on the other side can represent almost any desired policy but lack interpretability due to their black-box mapping.
In addition to that, if the NN cannot be pretrained \reviewed{using a reasonable behavior policy}, the NN parameters are usually initialized randomly \reviewed{requiring an end-to-end RL process}, leading to an excessive amount of required training data.
In this work, we focus on the approximation of the optimal policy via a parameterized MPC scheme.

MPC is a model-based optimal control approach that computes a sequence of optimal control actions~$\sequ^\star = ({\vu_0^\star}, \ldots, {\vu_{N-1}^\star})$ with $\vu_i \in \calA$ along a prediction horizon~$N \in \sN$ based on an internal system model such that a control objective is optimized and process constraint are satisfied.
\citet{grosDataDrivenEconomicNMPC2020} show that the following parameterized MPC optimization problem can be used as a policy in RL to capture the optimal policy 
\begin{subequations} \label{eq:MPC_OCP}
    \begin{align}
        \sequ^\star(\vs) = \arg \min_{\sequ} ~~& V_{\mathrm{f},\vtheta}(\vx_N) + \vw_\mathrm{f}^\top \vsigma_N \notag \\
        & +\sum_{i = 0}^{N-1} \ell_{\vtheta}(\vx_i, \vu_i) + \vw^\top \vsigma_i \label{eq:MPC_objective}\\ 
        \mathrm{s.t.} ~& \vx_{i+1} = f^{\mathrm{MPC}}_{\vtheta}(\vx_i, \vu_i),~ \vx_0 = \vs \label{eq:MPC_system}\\
        & \vu_\mathrm{lb} \leq \vu_i \leq \vu_{\mathrm{ub}} \label{eq:MPC_u_constraint}\\
        & g_{\vtheta}(\vx_i, \vu_i) \leq \vsigma_i,\quad \vsigma_i \geq 0 \label{eq:MPC_inequalities}\\
        & h_{\vtheta}(\vx_i, \vu_i) = 0 \label{eq:MPC_equalities} \\
        & g_{\mathrm{f}, \vtheta}(\vx_N) \leq \vsigma_N,\quad \vsigma_N \geq 0. \label{eq:MPC_terminal_const}
    \end{align}
\end{subequations}
In~\eqref{eq:MPC_OCP}, the objective function~\eqref{eq:MPC_objective} is minimized.
It consists of the sum of stage costs~$\ell_{\vtheta}:\calS \times \calA \rightarrow \sR$ starting from the initial condition~\reviewed{$\vs$} at $i = 0$ up to the second to last point of the prediction horizon at $i = N-1$, the terminal cost at the end of the prediction~$V_{\mathrm{f},\vtheta}:\calS \rightarrow \sR$, and penalties for soft constraints.
The system dynamics are accounted for via the system equations~$f^{\mathrm{MPC}}_{\vtheta}:\calS \times \calA \rightarrow \calS$ starting from the initial state~$\vx_0$ as stated in~\eqref{eq:MPC_system}.
The predicted trajectory does not need to be the same as in the MDP since $f^{\mathrm{MPC}}_{\vtheta}$ is usually only an approximation of the true system dynamics~\eqref{eq:MDP_transition_prob}.
Maximum and minimum values for the control action are posed via~\eqref{eq:MPC_u_constraint}.
Process equality and inequality constraints are ensured via~$h_{\vtheta}: \calS \times \calA \rightarrow \sR^{n_h}$ and $g_{\vtheta}: \calS \times \calA \rightarrow \sR^{n_g}$ as shown in~\eqref{eq:MPC_equalities} and~\eqref{eq:MPC_inequalities} respectively.
\reviewed{Terminal constraints~$g_{\mathrm{f},\vtheta}:\calS \rightarrow \sR^{n_{g_\mathrm{f}}}$ are represented via~\eqref{eq:MPC_terminal_const}.}
To always ensure feasibility, the inequality constraints \reviewed{and terminal constraints} need to be relaxed to soft constraints \reviewed{via slack variables~$\vsigma_i \in \sR^{n_g}$} \reviewed{and $\vsigma_N \in \sR^{n_{g_\mathrm{f}}}$}.
\reviewed{Also, $h_{\vtheta}$ needs to be designed such there always exists a feasible~$\vu_i$.}
All functions with the index~$\vtheta$ are parameterizable and can be adapted by the RL algorithm.
Note that to distinguish the states and actions within the optimal control problem~\eqref{eq:MPC_OCP} from those within the MDP, we introduce the states and actions within the MPC as~$\vx$ and~$\vu$.
After~\eqref{eq:MPC_OCP} is solved, the MPC policy is derived by taking the first element of the sequence of optimal control actions~$\sequ^\star$, so $\pi_{\vtheta}(\vs) = \vu_0^\star$. 

\subsubsection{Parametric NLP sensitivities}
In order to update the policy parameters with a policy optimization method, the Jacobian of the policy with respect to its parameters needs to be computed.
When using MPC as a policy, the derivative of the policy equals the derivative of the solution of the optimization problem, which is only computable explicitly in rare cases.
The MPC optimization problem~\eqref{eq:MPC_OCP} can be categorized as a parametric constrained nonlinear program~(NLP).
The derivative of the optimal solution with respect to the parameters of the NLP is also often referred to as the (parametric) NLP sensitivities.
We show next how the NLP sensitivities can be computed.

We first introduce a general constrained NLP of the form
\begin{subequations} \label{eq:generalNLP}
    \begin{align}
        \vz^\star(\vzeta) = \arg \min_{\vz} ~~& \Phi(\vz; \vzeta) \\
        \mathrm{s.t.}~~ & g(\vz; \vzeta) \leq 0 \\
        & h(\vz; \vzeta) = 0
    \end{align}
\end{subequations}
where $\vz\in \sR^{n_{\vz}}$ denotes the decision variables, $\vzeta \in \sR^{n_{\vzeta}}$ denotes the changeable parameters, $\Phi: \sR^{n_{\vz}} \times \sR^{n_{\vzeta}} \rightarrow \sR$ denotes the scalar objective function, $g:\sR^{n_{\vz}} \times \sR^{n_{\vzeta}} \rightarrow \sR^{n_g}$ denotes the inequality constraints, and $h: \sR^{n_{\vz}} \times \sR^{n_{\vzeta}} \rightarrow \sR^{n_h}$ denotes the equality constraints.
The optimal solution~$\vz^\star(\vzeta)$ of~\eqref{eq:generalNLP} changes with the optimization parameters~$\vzeta$.
This functional relationship is typically unknown.
We introduce the Lagrangian~$\calL: \sR^{n_{\vz}} \times \sR^{n_g} \times \sR^{n_h} \times \sR^{n_{\vzeta}} \rightarrow \sR$ of~\eqref{eq:generalNLP} as 
\begin{align}
    \calL(\vz, \vlambda, \vnu; \vzeta) = \Phi(\vz; \vzeta) + \vlambda^\top g(\vz;\vzeta) + \vnu^\top h(\vz;\vzeta)
\end{align}
where $\vlambda \in \sR^{n_g}$ and $\vnu\in \sR^{n_h}$ are the Lagrange multipliers, also known as dual variables, associated to the inequality and equality constraints respectively.
To simplify notation, we introduce the stacked primal-dual vector~$\vchi^\top = (\vz^\top, \vlambda^\top, \vnu^\top)$ with $\vchi \in \sR^{n_{\vchi}}$ and $n_{\vchi}= n_{\vz} + n_g + n_h$.
Let $\vchi^\star(\vzeta)$ denote an optimal primal-dual vector solving~\eqref{eq:generalNLP} and satisfying the Karush-Kuhn-Tucker~(KKT) conditions~\citep{nocedalNumericalOptimization2006}, and let $g$ satisfy the linear-independency constraint qualifications~\citep{nocedalNumericalOptimization2006}, then the equalities of the KKT conditions form an implicit function~$F:\sR^{n_{\vchi}} \times \sR^{n_{\vzeta}} \rightarrow \sR^{n_{\vchi}}$ of the form
\begin{align} \label{eq:NLP_implicit_function}
    F(\vchi^\star(\vzeta); \vzeta) = \begin{pmatrix}
    \nabla_{\vz} \calL(\vchi;\vzeta)\vert_{\vchi=\vchi^\star(\vzeta)} \\
    h(\vz^\star(\vzeta); \vzeta) \\
    \vlambda^\star(\vzeta) \odot g(\vz^\star(\vzeta);\vzeta)
    \end{pmatrix} = 0
\end{align}
where the operator~$\odot$ denotes the Hadamard (element-wise) product.
In order to obtain the first-order NLP sensitivities of the optimal primal-dual solution with respect to the NLP parameters~$\nabla_{\vzeta} \vchi^\star(\vzeta) \in \sR^{n_{\vchi}\times n_{\vzeta}}$,~\eqref{eq:NLP_implicit_function} can be implicitly differentiated almost everywhere~\citep{fiaccoSensitivityStabilityAnalysis1990} forming the following
linear system
\begin{align}
    \nabla_{\vchi^\star} F(\vchi^\star; \vzeta) \,\nabla_{\vzeta} \vchi^\star(\vzeta) = - \nabla_{\vzeta} F(\vchi^\star; \vzeta) \label{eq:IFT_for_sens}
\end{align}
with the coefficient matrix~$\nabla_{\vchi^\star} F(\vchi^\star; \vzeta) \in \sR^{n_{\vchi} \times n_{\vchi}}$ and the right-hand-side matrix~$-\nabla_{\vzeta} F(\vchi^\star;\vzeta) \in \sR^{n_{\vchi} \times n_{\vzeta}}$.
The first-order NLP sensitivities can therefore be obtained by solving~\eqref{eq:IFT_for_sens} for $\nabla_{\vzeta} \vchi^\star(\vzeta)$.
\reviewed{The second-order NLP sensitivities can be obtained in a similar fashion and also boil down to the solution of a linear system~\citep{brandnerReinforcedModelPredictive2024b, grosReinforcementLearningBased2021}}.
However the right-hand-side matrix requires third partial derivatives of the Lagrangian with respect to full primal-dual vector.
This can render the computation intractable, for instance if the number of primal-dual variables~$\vchi$ is large, or if highly nonlinear terms appear in the NLP such that the third partial derivatives become extremely convoluted.

To compute the deterministic policy gradient~\eqref{eq:DPG} and the respective terms of the deterministic policy Hessian~\eqref{eq:DPH_approx}, NLP~\eqref{eq:generalNLP} must be cast into~\eqref{eq:MPC_OCP} by setting~$\vz^\star = \sequ^\star$ and $\vzeta = \vtheta$.
The relevant sensitivities~$\nabla_{\vtheta} \pi_{\vtheta}(\vs) = \nabla_{\vtheta} \vu_{0, \vtheta}^\star(\vs)$ must then be extracted from the full sensitivity matrix.

\section{Robust computationally efficient Gauss-Newton policy optimization} \label{sec:Contribution}
Using MPC as policy within RL can significantly alleviate some of the main challenges of RL based on NN policies: it provides a good initial policy and can deal with hard constraints in real-time by solving an optimization problem.
Although the sparse parameterization of MPC policies enables the use of second-order optimization methods, the training still remains computationally challenging, as Hessian estimates such as~\eqref{eq:DPH_approx} rely on the second-order NLP sensitivities.
For that reason, we \reviewed{build on the findings in \citet{kordabadQuasiNewtonIterationDeterministic2022} by deriving an approximation for the Hessian matrix that avoids the computation of the second-order NLP sensitivities. We derive and propose a novel Gauss-Newton approximation of the deterministic policy Hessian that retains superlinear convergence while avoiding the computation of the second-order NLP sensitivities in Section~\ref{subsec:gauss-newton}.}
We then present a new approach to introduce momentum in the Gauss-Newton \reviewed{Hessian estimate} to increase numerical stability \reviewed{in Section~\ref{subsec:gauss-newton-momentum} before we combine both findings in the policy optimization algorithm with adaptive trust-region in Section~\ref{subsec:AdaptiveTR_PO}}.

\subsection{Gauss-Newton approximation of the deterministic policy Hessian}\label{subsec:gauss-newton}
In order to avoid the computation of the second-order NLP sensitivities, we will derive an approximation of the deterministic policy Hessian that does not depend on those, but still gives superlinear convergence guarantees.
Lets first consider the behavior of the action-value function of the optimal policy around the the optimal action~$\va = \pi^\star(\vs)$.
\begin{lemma} \label{th:Optimality_of_Q_func}
Let the action-value function of the optimal policy $Q^{\pi^\star}(\vs,\va)$ be differentiable in $\va$ at $\va = \pi_{\vtheta^\star}(\vs)$ for all $\vs\in\calS$, then the gradient evaluated at the optimal action is zero such that 
\begin{align}
    \nabla_{\va} Q^{\pi^\star} (\vs, \va) \vert_{\va = \pi^\star(\vs)} = 0, \quad \forall \vs \in \calS.
\end{align}
\end{lemma}
\begin{pf}
    Due to its definition, the optimal action~$\va = \pi^\star(\vs)$ has a better or at least equal value assigned to it when following the optimal policy~$\pi^\star$ than any other action, as it is defined in~\eqref{eq:optimal_action_value_function}.
    Therefore, $Q^{\pi^\star}(\vs,\pi^\star(\vs))$ is a maximum.
    Since $Q^{\pi^\star}(\vs, \va)$ is assumed to be continuously differentiable at $\va = \pi^\star(\vs)$, the first-order optimality condition is given as $\nabla_{\va} Q^{\pi^\star}(\vs, \va)\vert_{\va = \pi^\star(\vs)} = 0$. \qed   
\end{pf}
\begin{corollary}
    Let $Q^{\pi^\star}(\vs, \va)$ be twice differentiable in $\va$ at $\va = \pi^\star(\vs)$ then the Hessian with respect to $\va$ at $\va = \pi^\star(\vs)$ is negative semi-definite
    \begin{align}
        \nabla_{\va}^2 Q^{\pi^\star}(\vs, \va)\vert_{\va = \pi^\star(\vs)} \preceq 0.
    \end{align}
\end{corollary}
\begin{pf}
    To be a unique maximum, $\nabla_{\va}^2 Q^{\pi^\star}(\vs,\va)\vert_{\va= \pi^\star(\vs)}$ must be strictly negative definite.
    Since $Q^{\pi^\star} (\vs, \va)$ can also be locally constant for $\va$ in the neighborhood of $\pi^\star(\vs)$ if the optimal policy is not unique, the curvature can also be zero.
    Hence, $\nabla_{\va}^2 Q^{\pi^\star} (\vs, \va)\vert_{\va = \pi^\star(\vs)}$ is negative semi-definite. \qed
\end{pf}

\citet{kordabadQuasiNewtonIterationDeterministic2022} proves that the approximation of the deterministic policy Hessian stated in~\eqref{eq:DPH_approx} converges to the true deterministic policy Hessian for $\vtheta \rightarrow \vtheta^\star$.
\reviewed{The derived Hessian approximation obtains a superlinear convergence guarantee. However if MPC is used as a policy, it still requires to compute the second-order NLP sensitivities due to~\eqref{eq:DPH_part_1}, which can be computationally demanding. Building up on the findings in \citet{kordabadQuasiNewtonIterationDeterministic2022}, we} show that even when neglecting the first part~$\mM_1(\vtheta)$ of the approximation in~\eqref{eq:DPH_approx}, $\mM_2(\vtheta) \approx \nabla_{\vtheta}^2 J(\vtheta)$ converges to the true deterministic policy Hessian for $\vtheta \rightarrow \vtheta^\star$ with the benefit that the second-order NLP sensitivities can be omitted.
A Hessian approximation with the structure of~$\mM_2(\theta)$ is also referred to as a Gauss-Newton approximation~\citep{bottouOptimizationMethodsLargeScale2018}.
We first state the following assumption.
\begin{assumption} \label{ass:rich_policy}
The parameterized policy~$\pi_{\vtheta}(\vs)$ is rich e\-nough so there exists~$\vtheta^\star$ such that~$\pi_{\vtheta^\star}(\vs) = \pi^\star(\vs)$.
\end{assumption}
\begin{theorem} \label{th:mM_2_converges_to_H}
    Let Assumption~\ref{ass:rich_policy} hold and further let $\nabla_{\vtheta}^2 \pi_{\vtheta}(\vs)$ and $\nabla_{\va} Q^{\pi_{\vtheta}}(\vs, \va)\vert_{\va=\pi_{\vtheta}(\vs)}$ be bounded for all~$\vtheta$, then the approximation $\mM_2(\vtheta) \approx \nabla_{\vtheta}^2 J(\vtheta)$ converges for $\vtheta \rightarrow \vtheta^\star$ to the true deterministic policy Hessian at the optimal policy, so $\mM_2(\vtheta^\star) = \nabla_{\vtheta}^2 J(\vtheta)\vert_{\vtheta = \vtheta^\star}$ and moreover
    \begin{align}
        \lim_{\vtheta \rightarrow \vtheta^\star} \mM_1(\vtheta) = 0.
    \end{align}
\end{theorem}
\begin{pf}
    We define the auxiliary function~$\varphi_{\vtheta}:\calS\rightarrow \sR^{n_{\vtheta} \times n_{\vtheta}}$
    \begin{align}
        \varphi_{\vtheta}(\vs) = \nabla_{\vtheta}^2 \pi_{\vtheta}(\vs) \otimes \nabla_{\va} Q^{\pi_{\vtheta}}(\vs, \va)\vert_{\va = \pi_{\vtheta}(\vs)}.
    \end{align}
    Since~$\nabla_{\vtheta}^2\pi_{\vtheta}(\vs)$ and~$\nabla_{\va}Q^{\pi_{\vtheta}}(\vs,\va)\vert_{\va = \pi_{\vtheta}(\vs)}$ are bounded by assumption, $\varphi_{\vtheta}(\vs)$ is also bounded and can even be bounded uniformly.
    We compute the limit of $\varphi_{\vtheta}(\vs)$ for $\vtheta\rightarrow \vtheta^\star$ to be
    \begin{align}\label{eq:varphi_equal_0}
        &\lim_{\vtheta \rightarrow \vtheta^\star} \varphi_{\vtheta}(\vs) \notag \\
        = &\lim_{\vtheta \rightarrow \vtheta^\star} \nabla_{\vtheta}^2 \pi_{\vtheta}(\vs) \otimes \nabla_{\va} Q^{\pi_{\vtheta}}(\vs, \va)\vert_{\va = \pi_{\vtheta}(\vs)} \notag \\
        = &\lim_{\vtheta \rightarrow \vtheta^\star} \nabla_{\vtheta}^2 \pi_{\vtheta}(\vs) \otimes \lim_{\vtheta \rightarrow \vtheta^\star}\nabla_{\va} Q^{\pi_{\vtheta}}(\vs, \va)\vert_{\va = \pi_{\vtheta}(\vs)} \notag \\
        =& \nabla_{\vtheta}^2 \pi_{\vtheta^\star}(\vs) \otimes 0 = 0,
    \end{align}
    as it follows from \reviewed{Lemma}~\ref{th:Optimality_of_Q_func} that $\nabla_{\va}Q^{\pi_{\vtheta}}(\vs, \va)\vert_{\va=\pi_{\vtheta}(\vs)}$ vanishes at~$\vtheta^\star$ and $\nabla_{\vtheta}^2 \pi_{\vtheta}(\vs)$ is assumed to be bounded.
    
    From definition~\eqref{eq:DPH_part_1}, it follows that
    \begin{align}
        \lim_{\vtheta \rightarrow \vtheta^\star} \mM_1(\vtheta) &= \lim_{\vtheta \rightarrow \vtheta^\star} \sE_{\vs}\left[\varphi_{\vtheta}(\vs)\right] \notag \\
        &=\lim_{\vtheta \rightarrow \vtheta^\star} \int_{\calS} \rho_{\pi_{\vtheta}}(\vs)\varphi_{\vtheta}(\vs) \,\mathrm{d}\vs. \label{pr:integral}
    \end{align}   
    Here, $\rho_{\pi_{\vtheta}}(\vs):\calS \rightarrow [0, \infty)$ denotes the \reviewed{unnormalized} probability density that state~$\vs$ is encountered under closed-loop operation of~$\pi_{\vtheta}$.
    Since $\rho_{\pi_{\vtheta}}$ and $\varphi_{\vtheta}(\vs)$ can be uniformly bounded, the dominated convergence theorem~\citep{bronstejnTaschenbuchMathematik2016} can be applied so the limit operator can be moved inside the integral.
    It follows that
    \begin{align}
        &\lim_{\vtheta\rightarrow \vtheta^\star} \int_{\calS} \rho_{\pi_{\vtheta}}(\vs) \varphi_{\vtheta}(\vs) \, \mathrm{d}\vs \notag \\
        = &\int_{\calS} \lim_{\vtheta \rightarrow \vtheta^\star} \rho_{\pi_{\vtheta}}(\vs) \varphi_{\vtheta}(\vs) \, \mathrm{d}\vs \notag \\
        = &\int_{\calS} \lim_{\vtheta \rightarrow \vtheta^\star} \rho_{\pi_{\vtheta}}(\vs) \cdot \lim_{\vtheta \rightarrow \vtheta^\star} \varphi_{\vtheta}(\vs)  \, \mathrm{d}\vs \notag \\
        = &\int_{\calS} \rho_{\pi_{\vtheta^\star}}(\vs) \cdot 0 \, \mathrm{d} \vs = \int_{\calS} 0 \,  \mathrm{d}\vs = 0,
    \end{align}
    where the equality in the last line holds because of~\eqref{eq:varphi_equal_0}, proving that $\mM_1(\vtheta)$ vanishes if $\rho_{\pi_{\vtheta}} (\vs)$ is bounded. \qed
\end{pf}
\begin{corollary}
    Let Assumption~\ref{ass:rich_policy} hold and let $\mM_2(\vtheta)$ be defined for $\vtheta$ in the neighborhood of~$\vtheta^\star$, then $\mM_2(\vtheta)$ is negative semi-definite in the neighborhood of $\vtheta^\star$
    \begin{align}
        \mM_2(\vtheta) \preceq 0.
    \end{align}
\end{corollary}
\begin{pf}
    Since $\nabla_{\va}^2 Q^{\pi_{\vtheta}}(\vs, \va)\vert_{\va = \pi_{\vtheta}(\vs)}$ is negative semi-definite for $\vtheta$ in the neighborhood of $\vtheta^\star$, it follows from the fact that the result of a left and right multiplication with any matrix and its transposed to a negative semi-definite matrix is again a negative semi-definite matrix~\citep{bronstejnTaschenbuchMathematik2016} so
    \begin{align}
        \nabla_{\vtheta} \pi_{\vtheta}^\top(\vs) \nabla_{\va}^2 Q^{\pi_{\vtheta}}(\vs, \va)\vert_{\va = \pi_{\vtheta}(\vs)} \nabla_{\vtheta} \pi_{\vtheta}(\vs) \preceq 0.
    \end{align}
    As the expected value in the definition of $\mM_2(\vtheta)$ boils down to an integral over negative semi-definite matrices, the final result is also negative semi-definite. \qed
\end{pf}

Finally, we proof that choosing the approximate Hessian~$\mB_k$ as~$\mM_{2,k} = \mB_k$ still guarantees superlinear convergence.
For this, consider the condition for quasi-Newton methods for superlinear convergence.
\begin{theorem} \label{th:superlinear_convergence_of_QN_methods}
    Suppose that $J: \sR^{n_{\vtheta}} \rightarrow \sR$ is twice differentiable.
    Consider the iteration $\vtheta_{k+1} = \vtheta_k + \vp_k$ with $\vp_k$ defined as in~\eqref{eq:Quasi-Newton-Rule} and let the sequence of iterations $\{\vtheta_k\}$ converge to $\vtheta^\star$ such that $\nabla_{\vtheta} J(\vtheta)\vert_{\vtheta = \vtheta^\star} = 0$
    and $\nabla_{\vtheta}^2 J(\vtheta)\vert_{\vtheta = \vtheta^\star}$ is negative definite.
    Then $\{\vtheta_k\}$ converges superlinearly if 
    \begin{align}
        \lim_{k \rightarrow \infty} \frac{\| \left(\mB_k - \nabla_{\vtheta}^2 J(\vtheta)\vert_{\vtheta = \vtheta^\star}\right) \vp_k \|}{\|\vp_k\|} = 0. \label{eq:superlinearconvergence}
    \end{align}
\end{theorem}
\begin{pf}
    See Theorem 3.7. in \citet{nocedalNumericalOptimization2006}.
\end{pf}
\begin{corollary}
    Let Assumption~\ref{ass:rich_policy} hold, lets further assume that all assumptions from Theorem~\ref{th:superlinear_convergence_of_QN_methods} hold, and assume that the optimal policy is unique, then the sequence of iterations~$\{\vtheta_k\}$ in order to solve~\eqref{eq:parametric_policy_optimization} converges superlinearly to $\vtheta^\star$ if $\mB_k = \mM_{2,k}$ and update rule~\eqref{eq:Quasi-Newton-Rule} is used.
\end{corollary}
\begin{pf}
    Since a unique optimal policy is assumed, $\mM_2(\theta^\star)=\nabla_{\vtheta}^2J(\vtheta)\vert_{\vtheta = \vtheta^\star}$ is strictly negative definite.
    Further $\mB_k = \mM_{2,k}$ converges to $\nabla_{\vtheta}^2J(\vtheta)\vert_{\vtheta = \vtheta^\star}$ as shown in Theorem~\ref{th:mM_2_converges_to_H}.
    Therefore condition~\eqref{eq:superlinearconvergence} is satisfied. \qed
\end{pf}

\subsection{\reviewed{Robust Hessian estimation via momentum}}\label{subsec:gauss-newton-momentum}
If the action-value function~$Q^{\pi_{\vtheta}}(\vs, \va)$ of the current policy~$\pi_{\vtheta}(\vs)$ is explicitly known, and the expected values could be computed perfectly, the proposed approximate Newton method would converge superlinearly.
However, in reality, the action-value function of the current policy must be approximated. %
Further, the expected value must be computed empirically by drawing a finite amount of samples.
Both can lead to noisy estimates of the policy gradient~$\tilde{\vg}_k \approx \vg_k$ and the approximate policy Hessian~$\tilde{\mB}_k \approx \mB_k$ preventing accurate parameter updates and causing potential instability.
A common choice to mitigate the effect of noisy estimates within first-order stochastic optimization is the use of momentum-based approaches as introduced in Section~\ref{sec:first_order_optimization}.
While second-order updates are usually robust with respect to poorly estimated Hessian approximations~\citep{bollapragadaExactInexactSubsampled2019,bottouOptimizationMethodsLargeScale2018}, the estimates still need to be of reasonable accuracy.
A single poor Hessian estimation can already cause instability, for example if the eigenvalues that are estimated are orders of magnitude wrong, causing an update that deteriorates the parameter iterates to values that are far from optimal.
We show next how the ideas of momentum can be transferred to second-order approaches by applying some adaptations.

Analogously to the \reviewed{exponentially averaged gradient estimate~$\vm_k$, we define the biased exponentially} averaged Hessian estimate~$\mD_k\in\sR^{n_{\vtheta}\times n_{\vtheta}}$ recursively as
\begin{align}
    \mD_k = \eta \mD_{k-1} + (1 - \eta) \tilde{\mB}_k \label{eq:recursive_definition_D_k}
\end{align}
\reviewed{where} $0 \leq \eta < 1$ \reviewed{denotes} the exponential averaging factor for Hessian estimation, $\mD_{k-1}$ \reviewed{denotes} the biased averaged Hessian estimation of the previous iteration, and $\tilde{\mB}_k$ \reviewed{denotes} the local Hessian estimate at the current iteration.
Note that we do not specify~$\tilde{\mB}_k$ as it can be any Hessian approximation such as~$\tilde{\mB}_k = \tilde{\mM}_k$ or~$\tilde{\mB}_k = \tilde{\mM}_{2,k}$.

As in the case of the gradient momentum, the matrix~$\mD_0$ must be initialized.
Although it may be appealing to initialize with a zero matrix~$\mD_0 = 0$, it turns out to be impractical because especially in the first few steps of the optimization, $\mD_k$ would be severely influenced by inaccurate local estimates~$\tilde{\mB}_k$.
To avoid such undesirable behavior, we propose to initialize with
\reviewed{
the square root of the first estimate of the second momentum of the Adam optimizer, which is
\begin{align}
    \mD_0 = -\mathrm{diag}\left(\sqrt{\hat{\vv}_1} + \epsilon\right). %
\end{align}
}

To account for the initialization bias towards~$\mD_0$, we derive a scaling function similar to~\eqref{eq:unbiasing_of_gradient_estimation_w_momentum} that allows fast adjustments to newly obtained local Hessian estimates~$\tilde{\mB}_k$ and still maintains the added robustness from $\mD_0$.  
Writing the recursive definition~\eqref{eq:recursive_definition_D_k} explicitly and using
\reviewed{$-\mathrm{diag}\left(\sqrt{\hat{\vv}_1} + \epsilon\right) = (1- \eta)\tilde{\mB}_0$}
gives
\begin{align}
    \mD_k =& -\eta^k \reviewed{\mathrm{diag}\left(\sqrt{\hat{\vv}_1} + \epsilon\right)} + (1- \eta)\sum_{i = 1}^k \eta^{k-i} \tilde{\mB_i} \notag \\
    =&(1-\eta) \sum_{i = 0}^k \eta^{k-i} \tilde{\mB_i}.
\end{align}
This expression aligns closely to results from~\citet{kingmaAdamMethodStochastic2015} with the exception that the sum starts with $i=0$ instead of $i=1$.
Following a similar derivation as in~\citet{kingmaAdamMethodStochastic2015}, we derive the following corrected exponential moving average~$\hat{\mD}_k$
\begin{align}
    \hat{\mD}_k = \frac{\mD_k}{1 - \eta^{k+1}}. \label{eq:unbiased_D}
\end{align}
The derived expression differs from~\eqref{eq:unbiasing_of_gradient_estimation_w_momentum} by the shifted exponent.
This shift accounts for the fact that the initialization matrix~$\mD_0$ should be considered as the zeroth local Hessian estimate~$\tilde{\mB}_0$.
Lastly, since $0 \leq \eta < 1$, the denominator in~\eqref{eq:unbiased_D} converges to 1 such that
\begin{align}
    \hat{\mD}_{\infty} = \lim_{k\rightarrow\infty} \hat{\mD}_k =  \frac{\lim_{k\rightarrow\infty} \mD_k}{\lim_{k\rightarrow\infty} 1 - \eta^{k+1}} = \mD_{\infty}. \label{eq:biased_will_be_unbiased}
\end{align}

\begin{reviewedblock}
\subsection{Robust adaptive trust-region constrained policy optimization} \label{subsec:AdaptiveTR_PO}
Although second order approaches can lead to improved convergence behavior, they can also induce instabilities during training.
A frequently encountered scenario is that the Hessian of the objective function is singular, which is the case if the objective function behaves locally linear.
Even if regularization is applied to the Hessian, the computed (subsampled) Newton steps~\eqref{eq:Subsampled_Newton} can be very large.

Trust-region constrained optimization approaches account for such degenerate behavior.
They constrain the proposed update to be within a trust region radius~$\delta$ around the local iterate.
This trust region radius governs how conservative the local update is: a small trust region ensures stable but slow improvement, whereas a large trust region encourages large but potentially unstable steps.
Algorithms with an adaptive trust region $\delta = \delta_k$ aim at finding a compromise between updates with significant improvements but with a fair amount of stability.
A common approach is to compare the predicted improvement of the local quadratic approximation with the truly observed new objective value at the new iterate.

In RL, such comparative approaches are impractical due to the stochasticity of optimization problem~\eqref{eq:parametric_policy_optimization}.
\secondrevision{In fact, the expected value of the cumulative reward~\eqref{eq:Definition_CLC} needs to be estimated empirically.
As a consequence, the empirical estimate is noisy and would be compared to a deterministic model prediction.
If the empirical estimate is insufficiently accurate due to its noise, the trust region radius $\delta_k$ would decrease towards zero as the RL algorithm cannot find a feasible improvement although the model prediction may be good.}
Therefore, we propose to update the trust region radius~$\delta_k$ based on the 2-norm of the Adam update step~\eqref{eq:Adam_update} and embed it into the following trust-region constrained optimization problem
\begin{subequations}
    \begin{align}
        \vp_k = \arg \min_{\vp} \quad & \vp^\top \hat{\vm}_k + \frac{1}{2} \vp^\top \hat{\mD}_k \vp \\
        \mathrm{s.t.} \quad & \left\|\vp \right\|_2 \leq \alpha\left\|\frac{\hat{\vm}_k}{\sqrt{\hat{\vv}_k} + \epsilon}\right\|_2 = \delta_k .
    \end{align} \label{eq:Adaptive_TR_updates}
\end{subequations}
The proposed trust region constrained optimization problem with the adaptive trust region now allows to obtain large update steps if the gradients are large in magnitude and far away from the optimal solution, whereas the trust region is scaled down if the algorithm approaches a local optimum.
\end{reviewedblock}

\section{Analytical case study} \label{sec:Analytical_Case_Study}
We first show on an analytical example that the proposed Gauss-Newton approximation yields superlinear convergence in policy optimization.
Let us consider the following linear uncertain system
\begin{align}
    s^+ = s + a + w
\end{align}
where $w \sim \calN(0, \sigma_w^2)$ is a normal distributed uncertainty.
The system is controlled with the parameterized policy
\begin{align}
    \pi_{\theta}(s) = - \theta^2 s,
\end{align}
where the coefficient is chosen as $\theta^2$ instead of~$\theta$ to ensure that the approximate Hessian~\eqref{eq:DPH_approx} and the Gauss-Newton approximation yield different results due to~$\nabla_{\theta}^2\pi_{\theta}(s) \neq 0$.
The first and second-order policy derivatives are
\begin{align}
    \nabla_{\theta}\pi_\theta (s) = -2 \theta s, \quad \nabla_{\theta}^2 \pi_\theta(s) = -2s.
\end{align}
We consider the reward function~$r(s,a) = -0.5(s^2 + a^2)$.
One can show that the state-value function~$V^{\pi_\theta}(s)$ for the closed-loop system with $0 < \gamma < 1$ has the following structure
\begin{align}
    V^{\pi_{\theta}}(s) = p_{\theta} s^2 + q_{\theta}
\end{align}
where $p_\theta$ and $q_\theta$ are coefficients that depend on~$\theta$.
From the definition of state-value function~\eqref{eq:state_value_function_recursively} it follows that
\begin{align}
    V^{\pi_\theta}(s) 
    =& -0.5 \left(s^2 + (-\theta^2 s)^2\right) \notag \\
    &+ \gamma \sE_{w \sim \calN(0, \sigma_w^2)} \left[ V^{\pi_\theta}(s - \theta^2 s + w)\right] \notag \\
    =& \left[-0.5\left(1 + \theta^4\right) + \gamma \left(1 - \theta^2 \right)^2 p_\theta\right]s^2 \notag \\
    &+ \gamma \left(q_\theta + p_\theta \sigma_w^2\right).
\end{align}
From a comparison of coefficients, the following expressions for $p_\theta$ and $q_\theta$ can be derived
\begin{align}
    p_{\theta}= -\frac{0.5 \left(1 + \theta^4\right)}{1 - \gamma \left(1 - \theta^2\right)^2}, \quad q_{\theta} = \frac{\gamma}{1 - \gamma} \sigma_{w}^2 p_{\theta}. \label{eq:Analytical_pq}
\end{align}
The action-value function~$Q^{\pi_\theta}(s,a)$ can be written as
\begin{subequations}
    \begin{align}
        Q^{\pi_\theta}(s,a) =& -0.5\left(s^2 + a^2\right) \notag \\
        &+ \gamma \sE_{w \sim \calN(0, \sigma_w^2)} \left[V^{\pi_\theta}(s + a + w) \right] \\
        =& -0.5\left(s^2 + a^2\right) + \gamma p_\theta (s + a)^2 \notag \\
        &+ \gamma \left(q_\theta + p_\theta \sigma_w^2\right)
    \end{align}
\end{subequations}
with gradient and Hessian with respect to $a$ given as 
\begin{subequations}
    \begin{align}
        \nabla_a Q^{\pi_\theta}(s,a) =& (2\gamma p_\theta - 1) a + 2 \gamma p_\theta s, \\
        \nabla_a^2 Q^{\pi_\theta}(s,a) =& 2\gamma p_\theta - 1.
    \end{align}
\end{subequations}

Let the initial state be normally distributed~$s_0 \sim \calN(0, \sigma_0^2)$, then the expected cumulative reward~$J(\theta)$ can be derived from its definition~\eqref{eq:Definition_CLC} as
\begin{align}
    J(\theta) = \sE_{s_0\sim\calN(0,\sigma_0^2)}\left[V^{\pi_\theta}(s_0)\right] = q_\theta + p_\theta \sigma_0^2.
\end{align}
The exact gradient and the Hessian with respect to $\theta$ can therefore be derived as
\begin{subequations}
    \begin{align}
        \nabla_\theta J(\theta) =& \nabla_\theta q_\theta + \sigma_0^2 \nabla_{\theta} p_\theta, \label{eq:LQR_DPG1}\\
        \nabla_\theta^2 J(\theta) =&  \nabla^2_\theta q_\theta + \sigma_0^2 \nabla^2_{\theta} p_\theta. \label{eq:LQR_DPH}
    \end{align}
\end{subequations}
Setting~\eqref{eq:LQR_DPG1} to zero and substituting~\eqref{eq:Analytical_pq} allows to derive the optimal parameter~$\theta^\star(\gamma)$ as 
\begin{align}
    \theta^\star(\gamma) = \sqrt{\frac{\sqrt{1 + 4 \gamma^2} - 1}{2\gamma}},
\end{align}
which computes to $\theta^\star \approx 0.767$ for $\gamma = 0.9$.

Next, we will derive the expressions for the Gauss-Newton approximation and the Hessian approximation according to~\eqref{eq:DPH_approx}. 
Since the policy gradient~$\nabla_{\theta}J(\theta)$ can also be obtained from~\eqref{eq:DPG} we have
\begin{align}
    \nabla_\theta J(\theta) =& \sE_{s}\left[(-2\theta s) \cdot\left((2 \gamma p_\theta - 1)(-\theta^2s) +2 \gamma p_\theta s\right)\right] \notag \\
    =& \left(2\theta^3 (2 \gamma p_\theta -1) - 4 \gamma p_\theta \theta\right) \sE_s\left[s^2\right], \label{eq:LQR_DPG2}
\end{align}
which gives the following expression for~$\sE_{s}\left[s^2\right]$ after equating~\eqref{eq:LQR_DPG1} and~\eqref{eq:LQR_DPG2}
\begin{align}
    \sE_{s}\left[s^2\right] = \frac{\nabla_{\theta}q_\theta + \sigma_0^2 \nabla_{\theta}p_\theta}{2\theta^3 (2 \gamma p_\theta -1) - 4 \gamma p_\theta \theta}.
\end{align}
We derive the matrices $M_1(\theta)$ and $M_2(\theta)$ from the definitions~\eqref{eq:DPH_part_1} and~\eqref{eq:DPH_part_2} as
\begin{subequations}
    \begin{align}
        M_1(\theta) =& 2 \theta^2(2 \gamma p_\theta - 1) \sE_s \left[s^2\right] - 4 \gamma p_\theta \sE_s \left[s^2\right] \label{eq:LQR_DPH_approx_1}\\
        M_2(\theta) =& 4 \theta^2 (2 \gamma p_\theta - 1) \sE_{s}\left[s^2\right]. \label{eq:LQR_DPH_approx_2}
    \end{align}
\end{subequations}

\begin{figure}
    \centering
    \includegraphics[width=\linewidth]{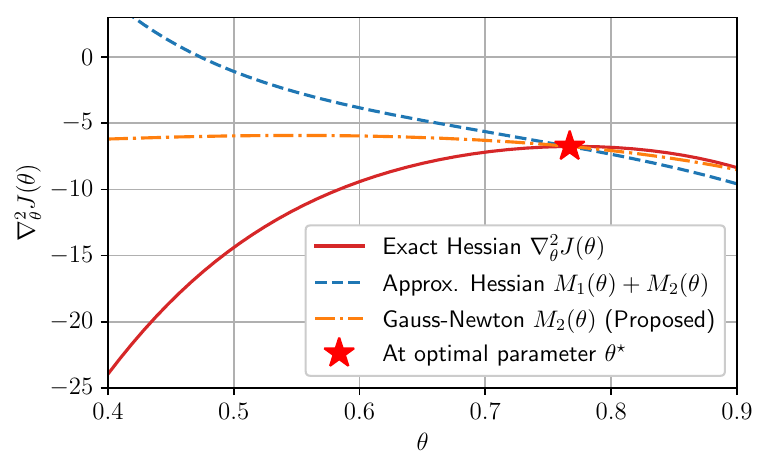}
    \caption{Hessian and its approximations over~$\theta$ for $\gamma = 0.9$ and $\sigma_w^2 = \sigma_0^2 = 0.1$. The Hessian at~$\theta^\star$ is highlighted as a star.}
    \label{fig:LQR_Hessians}
\end{figure}

Figure~\ref{fig:LQR_Hessians} shows the exact Hessian~\eqref{eq:LQR_DPH}, the approximate Hessian as the sum of~\eqref{eq:LQR_DPH_approx_1} and~\eqref{eq:LQR_DPH_approx_2}, and the proposed Gauss-Newton approximation~\eqref{eq:LQR_DPH_approx_2} for a discount factor of~$\gamma = 0.9$ under variation of~$\theta$.
All curves intersect at~$\theta^\star$. %
Further, the proposed Gauss-Newton approximation remains negative definite over the considered range whereas the more elaborate approximation becomes positive definite far from the optimal parameter. 

\begin{figure}
    \centering
    \includegraphics[width = \linewidth]{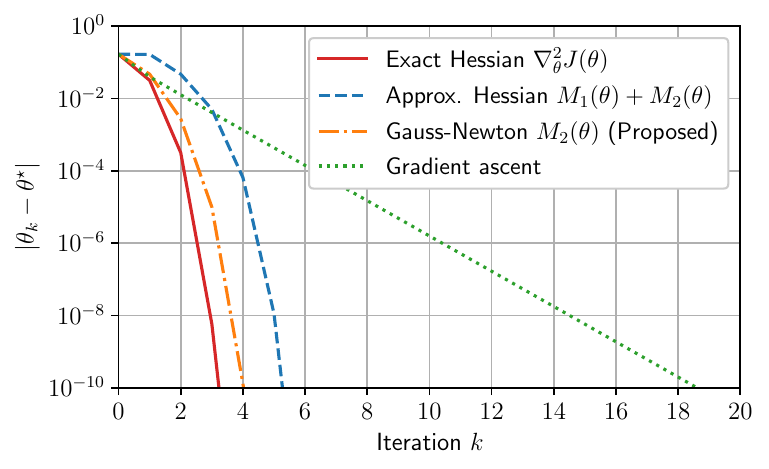}
    \caption{Error between the current iterate and the optimal parameter~$| \theta_k -\theta^\star|$ for the second-order approaches and first-order gradient ascent starting at the initial guess~$\theta_0 = 0.6$.}
    \label{fig:LQR_superlinear_convergence}
\end{figure}

Lastly, we investigate the error dynamics~$|\theta_k - \theta^\star|$ between the current iterate~$\theta_k$ and the optimal parameter~$\theta^\star$ for the considered second-order approaches and gradient ascent.
Figure~\ref{fig:LQR_superlinear_convergence} shows the error dynamics for an initial guess of~$\theta_0 = 0.6$.
All second-order approaches use a learning rate of~$\alpha = 1$ whereas the gradient ascent approach uses a learning rate of~$\alpha = 0.1$.
The second-order approaches show a superlinear convergence rate.
The first-order gradient ascent algorithm only achieves a linear error reduction.

\section{Continuously stirred tank reactor}  \label{sec:CSTR}
We demonstrate \reviewed{the scalability with respect to the number of parameters} of the proposed \reviewed{adaptive trust region constrained} Gauss-Newton approach in comparison to more elaborate but more costly approximate Hessian methods like~\eqref{eq:DPH_approx} at the example of a continuously stirred tank reactor~(CSTR).
\reviewed{
We also investigate the resilience of the proposed adaptive trust region constrained Gauss-Newton approach with respect to parameter scales.
All investigations are compared to the Adam optimizer~\citep{kingmaAdamMethodStochastic2015}.
Lastly, the proposed approach is compared to deep RL policies trained with the TD3 algorithm~\citep{fujimotoAddressingFunctionApproximation2018} with respect to learning efficiency and final closed-loop performance. 
}  
All code is available online\footnote{https://github.com/DeanBrandner/GaussNewtonRLforMPC}.

\subsection{System representation}
\begin{figure}
    \centering
    \includegraphics[width =0.8\linewidth]{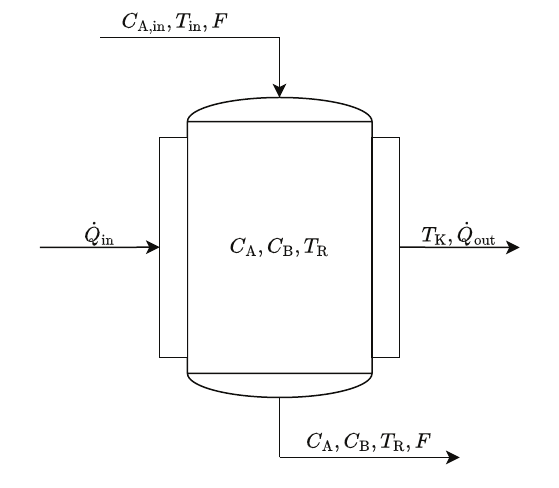}
    \caption{CSTR with entering and leaving streams.} \label{fig:CSTR}
\end{figure}

The investigated CSTR is taken from~\citet{klattGainschedulingTrajectoryControl1998}.
We consider the following reaction scheme
\begin{align}
    \mathrm{A} \stackrel{k_1}{\longrightarrow} \mathrm{B} \stackrel{k_2}{\longrightarrow} \mathrm{C}, \quad 2\mathrm{A} \stackrel{k_3}{\longrightarrow} \mathrm{D}
\end{align}
of educt $\mathrm{A}$ to value product $\mathrm{B}$ with the consecutive reaction of $\mathrm{B}$ to side product $\mathrm{C}$ and the parallel reaction of $\mathrm{A}$ to side product $\mathrm{D}$.
The reaction constants $k_i$ for each reaction are placed above the respective arrows. 
Figure~\ref{fig:CSTR} illustrates the streams entering and leaving the system.
A stream with a dilution rate of~$F$, a concentration of A~$C_\mathrm{A,in}$, and a temperature of~$T_\mathrm{in}$ enters the reactor, and leaves it with the same dilution rate~$F$ but concentrations~$C_\mathrm{A}$, $C_\mathrm{B}$, and temperature~$T_\mathrm{R}$.  
The CSTR consists of a cooling jacket with an entering heat stream of $\dot{Q}_\mathrm{in}$ and a leaving heat stream of~$\dot{Q}_\mathrm{out}$ with temperature~$T_\mathrm{K}$.
The difference between the entering and leaving heat streams is denoted as~$\dot{Q}= \dot{Q}_\mathrm{in} - \dot{Q}_\mathrm{out}$ and forms together with the dilution rate the manipulated variables~$\va^\top = \vu^\top = (F, \dot{Q})$.
The physical state of the system is formed as~$\vs_{\mathrm{phys}}^\top = \vx^\top = (C_\mathrm{A}, C_\mathrm{B}, T_\mathrm{R}, T_\mathrm{K})$.
It  is governed by the following ODEs
\begin{subequations} \label{eq:CSTR_system_equations}
    \begin{align}
        \dot{C}_{\mathrm{A}} &= F (C_{\mathrm{A,in}} - C_{\mathrm{A}}) - k_1  C_{\mathrm{A}} - k_3 C_{\mathrm{A}}^2, \\
        \dot{C}_{\mathrm{B}} &= -F C_{\mathrm{B}} + k_1 C_{\mathrm{A}} - k_2 \, C_{\mathrm{B}}, \\
        \dot{T}_{\mathrm{R}} &= \frac{k_1  C_{\mathrm{A}}  H_{\mathrm{R, ab}} + k_2  C_{\mathrm{B}}  H_{\mathrm{R,bc}} + k_3  C_{\mathrm{A}}^2  H_{\mathrm{R,ad}}} {-\varrho  c_p} \notag\\
        &+ F  (T_{\mathrm{in}} - T_{\mathrm{R}}) + \frac{K_w \, A_{\mathrm{R}} \,(T_{\mathrm{K}}-T_{\mathrm{R}})}{\varrho \, c_p \, V_{\mathrm{R}}}, \\
        \dot{T}_{\mathrm{K}} &= \frac{\dot{Q} + K_w \, A_{\mathrm{R}} \left( T_{\mathrm{R}} - T_\mathrm{K} \right)}{m_k  c_{p,\mathrm{K}}},
    \end{align}
\end{subequations}
where the reaction rates are defines as
\begin{subequations} \label{eq:CSTR_reaction_rates}
    \begin{align}
        k_1 &= \mu_{\beta} k_{0,\mathrm{ab}}  \exp\left(\frac{-E_{\mathrm{A},\mathrm{ab}}}{R \,(T_{\mathrm{R}}+273.15)}\right), \label{subeq:CSTR_kinetic_1} \\
        k_2 &= k_{0,\mathrm{bc}}  \exp \left( \frac{-E_{\mathrm{A},\mathrm{bc}}}{R\,(T_{\mathrm{R}}+273.15)} \right), \label{subeq:CSTR_kinetic_2}\\
        k_3 &= k_{0,\mathrm{ad}}  \exp \left( \frac{-\mu_{\alpha} E_{\mathrm{A},\mathrm{ad}}}{R \, (T_{\mathrm{R}}+273.15)} \right). \label{subeq:CSTR_kinetic_3}
    \end{align}
\end{subequations}
The parameters are defined as the nominal values in~\citet{klattGainschedulingTrajectoryControl1998}.
The coefficients~$\mu_\alpha$ and $\mu_\beta$ will be uncertain in the following investigations.

\subsection{Agent-environment interaction}
The goal is to control the CSTR to a reactor temperature of $T_\mathrm{R,ref} = 126\,^{\circ}\mathrm{C}$ and a jacket temperature of~$T_\mathrm{K,ref} = 120\,^{\circ}\mathrm{C}$.
Both setpoints are obtained by robust steady state optimization.
Further, the variations in control input should be kept low to achieve smooth control trajectories.
To account for that, the following algebraic equations for the previous control inputs must be introduced
\begin{align}
    F_\mathrm{prev} = F, \quad \dot{Q}_\mathrm{prev} = \dot{Q}.
\end{align}
They form the state of previous control inputs as~$\vs_\mathrm{prev}^\top = (F_\mathrm{prev}, \dot{Q}_\mathrm{prev})$ and complete the definition of the RL state~$\vs^\top = (\vs_\mathrm{phys}^\top, \vs_\mathrm{prev}^\top)$.
The states and actions also need to satisfy constraints~ $\vs_\mathrm{lb} \leq \vs \leq \vs_\mathrm{ub}$ and $\va_\mathrm{lb} \leq \va \leq \va_\mathrm{ub}$.
The upper and lower bounds are listed in Table~\ref{tab:Upper_and_lower_bounds}.
We define the reward~$R(\vs,\va)$ that encodes the objective above as
\begin{subequations}
    \begin{align}
        R(\vs, \va) = & R_\mathrm{ref}(\vs,\va) + R_{\Delta \va}(\vs, \va) + R_{\vsigma}(\vs, \va) \label{eq:CSTR_total_reward}\\
        R_\mathrm{ref}(\vs, \va) = &-10^{-2} \cdot (T_\mathrm{R} - T_\mathrm{R,ref})^2 \notag \\ 
         &- 10^{-2} \cdot(T_\mathrm{K} - T_\mathrm{K,ref})^2 \label{eq:CSTR_ref_reward} \\
        R_{\Delta \va}(\vs, \va) = &-10^{\reviewed{2}} \cdot \left(\frac{F_\mathrm{prev} - F}{F_\mathrm{max} -F_\mathrm{min}} \right)^2 \notag \\
        &- 10^{\reviewed{2}} \cdot \left(\frac{\dot{Q}_\mathrm{prev} - \dot{Q}}{\dot{Q}_\mathrm{max} - \dot{Q}_\mathrm{min}} \right)^2 \\
         R_{\vsigma}(\vs, \va) = &-10^2 \max \left\{0, \vs_\mathrm{lb} - \vs, \vs - \vs_\mathrm{ub}\right\} \label{eq:CSTR_penalty_reward}
    \end{align}
\end{subequations}
where $R_\mathrm{ref}(\vs, \va)$ encodes the tracking of the references, $R_{\Delta \va}(\vs, \va)$ encodes the low variations in the actions, and $R_{\vsigma}(\vs, \va)$ encodes the constraints \secondrevision{as soft constraints}.
\secondrevision{Note that a soft constraint formulation does not fully prohibit constraint violations. Slight constraint violations may still be possible even for the optimal policy.}

\begin{table}
    \centering
    \caption{Upper and lower bounds\,(ub, lb) for states~$\vs$ (left) and actions~$\va$ (right). Concentrations~$C_i$ are given in $\mathrm{mol} \,\mathrm{L}^{-1}$, temperatures~$T_i$ in~$^\circ \mathrm{C}$, dilution rates~$F_i$ in~$\mathrm{h}^{-1}$, and heat flows~$\dot{Q}_i$ in~$\mathrm{kW}$.} \label{tab:Upper_and_lower_bounds}
    \begin{tabular}{ccccccc|cc} \toprule
    & $C_\mathrm{A}$ & $C_\mathrm{B}$ & $T_\mathrm{R}$ & $T_\mathrm{K}$ & $F_\mathrm{prev}$ & $\dot{Q}_\mathrm{prev}$ & $F$ & $\dot{Q}$ \\ \midrule \midrule
         lb &  $0.1$ & $0.1$ & $80$ & $80$ & $5$ & $-8.5$ & $5$& $-8.5$ \\
         ub &  $2$ & $2$ & $140$ & $140$ & $40$ & $0$ & $40$ & $0$ \\ \bottomrule
    \end{tabular}
\end{table}

The initial states~$\vs_0$ of the environment from which all episodes start are sampled uniformly from the feasible state space, so~$\vs_0 \sim \calU(\vs_\mathrm{lb}, \vs_\mathrm{ub})$.
Further, it is assumed that the coefficients~$\mu_\alpha$ and $\mu_\beta$ \reviewed{in \eqref{subeq:CSTR_kinetic_3} and \eqref{subeq:CSTR_kinetic_1}} are uncertain.
Both coefficients are sampled from the uniform distributions~$\mu_\alpha \sim \calU(0.95, 1.05)$ and ~$\mu_\beta \sim \calU(0.95, 1.05)$ \reviewed{for each transition. Additionally, the subsequent physical state $\vs^+_\mathrm{phys}$ is disturbed with additive process noise~$\vw\sim\calN(\cdot\vert \mu = 0, \mSigma = \mathrm{diag}(0.025, 0.025, 0.5, 0.5))$.}
The sampled realizations of $\mu_\alpha$ and $\mu_\beta$ \reviewed{as well as the realization of the additive uncertainty~$\vw$} are unknown to the agent, which means that the agent needs to learn a control policy that maximizes the cumulative reward but remains robust simultaneously.

The considered agent is a nominal MPC agent that uses the same model as described in~\eqref{eq:CSTR_system_equations} and~\eqref{eq:CSTR_reaction_rates}, respects the constraints from Table~\ref{tab:Upper_and_lower_bounds}, takes the negative reward from~\eqref{eq:CSTR_total_reward} as stage cost, and predicts $N = 20$ steps into the future.
\reviewed{
The MPC parameterization will vary in the subsequent subsections.
We want to emphasize that the MPC policy does not necessarily need to use the exact system model to predict the future.
Practical alternatives such as data-driven surrogate models or linear systems can also be used and parameterized.
}
\reviewed{
In order to investigate the effect of the proposed adaptive trust region constrained Gauss-Newton approach, we avoid to approximate the Q-function with the help of function approximators as these can significantly affect computation time and introduce other undesired side effects.
We therefore implement the system, the reward and the MPC in a differentiable fashion such that the gradient and Hessian of the Q-function can be computed.
The proposed approach is not constrained to differentiable models but can also be applied to the fully model-free setting.
As with conventional first order approaches, an accurate approximation of the Q-function, optimally using a compatible function approximator as derived in works like~\cite{silverDeterministicPolicyGradient2014} and \cite{kordabadQuasiNewtonCompatibleActorCritic2025}, should be used if possible to avoid any systematic errors.
}
All main training routines are based on the python toolbox do-mpc~\citep{fiedlerDompcFAIRNonlinear2023}, which uses a CasADi~\citep{anderssonCasADiSoftwareFramework2019} backend and relies on Ipopt~\citep{wachterImplementationInteriorpointFilter2006}.

\begin{reviewedblock}
\subsection{Scalability with the number of parameters} \label{subsec:CSTR_Scalability}
We first investigate the scalability of the proposed adaptive trust region constrained Gauss-Newton approach with respect to the number of trainable parameters.
We train three increasingly more parameterized MPC agents, ranging from a low parameterized agent with two parameters, over a medium parameterized agent with 13 parameters, up to a highly parameterized agent with 33 parameters.
The low parameterization reflects a high level of certainty in almost all system parameters, whereas a high parameterization reflects as low level of certainty, allowing the RL algorithm to tune the parameters to come up with an optimized policy.
The low parameterized agent considers the uncertain parameters \secondrevision{$E_{A,\mathrm{ad}}$} and \secondrevision{$k_{0,\mathrm{ab}}$} to be trainable, whereas the medium parameterized agent also considers other kinetic \secondrevision{and thermodynamic} parameters\secondrevision{, and the setpoints within the objective} to be trainable.
The highly parameterized agent also parametrizes the terminal cost.
\secondrevision{Each introduced parameterization shares a similar structure of 
\begin{align}
\tilde{\vp}_{i,\vtheta} = (1 + \vtheta_i) c_i \vp_i \label{eq:systematic_parameterization}
\end{align}
where $\tilde{\vp}_i$ denotes the $i$-th MPC parameter that is modified, $\vp_i$ denotes the $i$-th nominal MPC parameter, $c_i$ denotes a coefficient that models model error, and $\vtheta_i$ is the $i$-th trainable policy parameter.
The terminal cost~$V_{\mathrm{f},\vtheta}$ is learned according to the following parameterization
\begin{align}
    V_{\mathrm{f},\vtheta} = \left(\vx_N - \tilde{\vx}_{\mathrm{ref,f}, \vtheta}\right)^\top \tilde{\mP}_{\vtheta}^\top \tilde{\mP}_{\vtheta}  \left(\vx_N - \tilde{\vx}_{\mathrm{ref,f}, \vtheta}\right)
\end{align}
with cost matrix $\tilde{\mP}_{\vtheta} = \mathrm{diag}\left(0, 0, 10^{-2}, 10^{-2}\right) + \mP_{\vtheta}$, where $\mP_{\vtheta} \in \sR^{4 \times 4}$ is a matrix with RL parameters in each cell, and setpoint $\vx_\mathrm{ref,f}^\top = (0.5, 0.5, 126, 120)$.
Detailed information on the parameterization levels is provided in Table~\ref{tab:Parameterizations}.
}
\begin{secondrevisionblock}
\begin{table}[]
    \centering
    \caption{Parameters for the different parameterization levels.}
    \label{tab:Parameterizations}
   \begin{tabularx}{\linewidth}{C|CCC|CCC}

        \toprule
         Level & $i$ & $\vp_i$ & $c_i$ & $i$ & $\vp_i$ & $c_i$ \\
        \midrule\midrule

        \multirow{1}{*}{Low}
        & $1$ 
        & $E_{A,\mathrm{ad}}$
        & $0.95$ & $2$ & $k_{0,\mathrm{ab}}$ & $1$\\
        \midrule

        \multirow{6}{*}{Med}
        & $3$ & $k_{0,\mathrm{bc}}$ & $1$
        & $4$ & ${k}_{0,\mathrm{ad}}$ & $1$ \\

        & $5$ & $E_{A,\mathrm{ab}}$ & $0.95$
        & $6$ & $E_{A,\mathrm{bc}}$ & $1.05$ \\

        & $7$ & $H_{\mathrm{R,ab}}$ & $1.05$
        & $8$ & $H_{\mathrm{R,bc}}$ & $1.05$ \\

        & $9$ & $H_{\mathrm{R,ad}}$ & $0.95$
        & $10$ & $c_p$ & $0.98$ \\

        & $11$ & $c_{p,\mathrm{K}}$ & $1.02$
        & $12$ & $T_{\mathrm{R,ref}}$ & $1.0$ \\

        & $13$ & $T_{\mathrm{K,ref}}$ & $1$
        & $-$ & $-$ & $-$ \\
        \midrule

        \multirow{1}{*}{High}
        & $14 \ldots29$ & $\mP_{\vtheta}$ & $1$ & $30\ldots 33$& $\vx_{\mathrm{ref,f}}$ & $1$ \\
        \bottomrule
    \end{tabularx}
\end{table}
\end{secondrevisionblock}

Each agent is trained with the first order Adam optimizer, the adaptive trust region constrained second order approach with Hessian approximation according to~\eqref{eq:DPH_approx}, and the proposed adaptive trust region constrained Gauss-Newton approach.
The Adam optimizer uses the default hyperparameters of $\beta_1 = 0.9$ and $\beta_2 = 0.999$ with learning rate $\alpha=10^{-2}$.
Larger learning rates lead to instabilities.
Both second order approaches share the same hyperparameters of $\beta_1 = 0.75$, $\beta_2 = 0.999$, $\eta = 0.9$ and $\alpha = 0.1$.
For each RL iteration~$k$, the agents collect data from $N_\mathrm{IC} = 200$ initial conditions with a fixed length of $N_{\mathrm{ep},i} = 100$ steps.

Figure~\ref{fig:LearningCurvesScalability} shows the learning curves for the three differently parameterized MPC agents trained with the three approaches.
All approaches show stable training.
The proposed Gauss-Newton approach improves more rapidly than the Adam optimizer especially in the first iterations, showing that the Gauss-Newton approach can lead to an improved performance with less data.
Simultaneously, the approximate Newton approach also shows stable improvements, however it shows mixed results regarding convergence speed.
While it turns out to be faster than the Adam optimizer for the low and high parametrization, it improves slower in the scenario with medium parameterization.
We account this to the fact that inaccurate intermediate Hessian estimates can adversely affect high quality updates.
This deterioration then first needs to be recovered from, compromising the improved convergence speed.
\begin{figure}
    \centering
    \includegraphics[width=\linewidth]{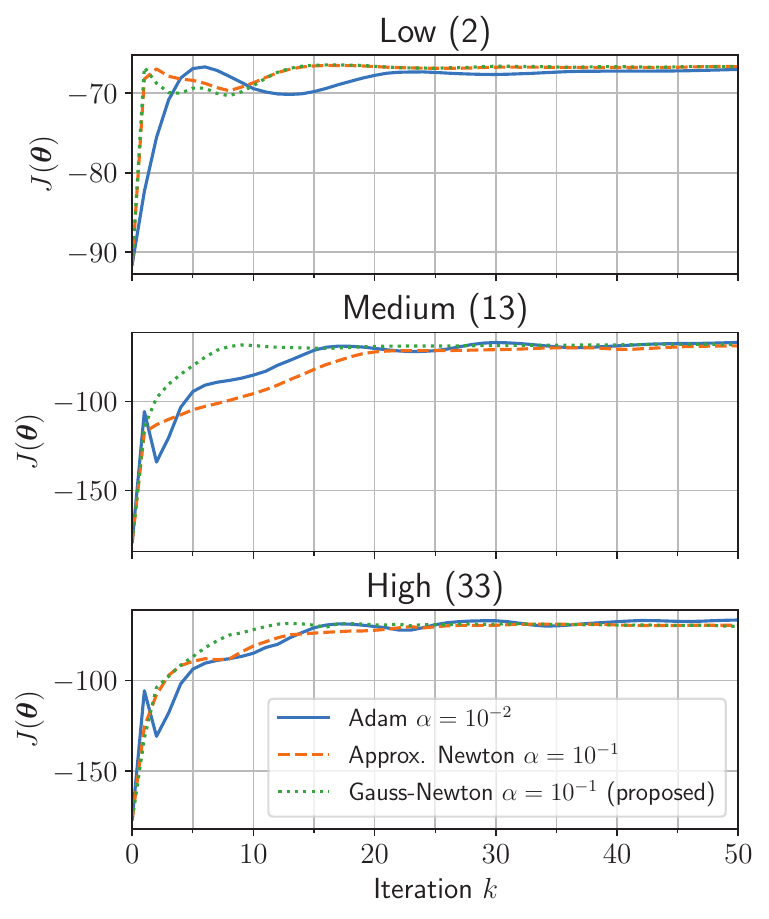}
    \caption{Expected cumulative reward for increasingly more parameterized MPC schemes. The number in parentheses denotes the total number of trainable parameters.}
    \label{fig:LearningCurvesScalability}
\end{figure}

\begin{figure}
    \centering
    \includegraphics[width=\linewidth]{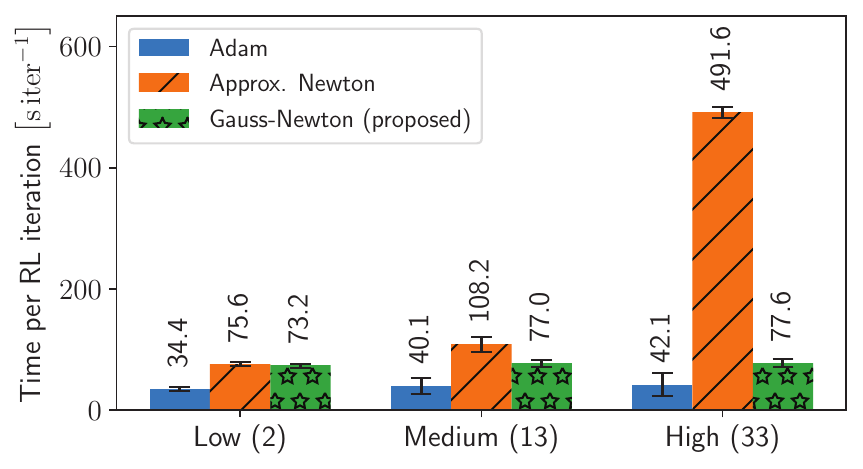}
    \caption{Mean time per RL iteration for differently parameterized MPC schemes.
    \secondrevision{The numbers above each bar show the mean value.}
    The error bars denote three standard deviations.
    The number in parentheses denotes the total number of trainable parameters.}
    \label{fig:ComputationTimesScalability}
\end{figure}
Figure~\ref{fig:ComputationTimesScalability} shows the average time per RL iteration for the investigated scenarios.
For all three parameterizations, the Adam optimizer has the smallest iteration time, followed by the proposed Gauss-Newton approach, which takes about double the iteration time in the investigated scenario.
In both approaches, the iteration time increases moderately with an increasing number of parameters.
This stays in contrast to the approximate Newton approach, for which the iteration time increases drastically.
We account the drastic increase of the iteration time to the fact that in order to compute the Hessian approximation according to~\eqref{eq:DPH_approx}, the second order NLP sensitivities need to be computed, which scales quadratically with the number of parameters.
This is not the case for the proposed Gauss-Newton approximation.

We therefore conclude that the proposed Gauss-Newton approach offers a computationally cheap alternative to second order approaches with Hessian approximation that use second order NLP sensitivities such as~\eqref{eq:DPH_approx} since the computation time scales better to higher parameterized policies.
In addition to that, we see that the computation time of the Gauss-Newton approach scales similar to the computation time of the Adam optimizer.
Although the iteration cost of the Gauss-Newton approach is higher in the presented case study, the number of iterations until a reasonable performance is achieved is less.
This renders the proposed Gauss-Newton approach of special interest for settings in which the number of interactions between the agent and the environment are the limiting factor.
\end{reviewedblock}

\begin{reviewedblock}
\subsection{Influence of the parameter scaling}
Next, we investigate how the curvature with respect to the parameters affects the learning stability.
This is of high interest since it is typically difficult to assess the sensitivity of the objective with respect to the policy parameters and its rate of change beforehand.
Moreover especially for MPC policies, parameters can have significantly different influences.
This affects first order optimization approaches in particular, as the maximum learning rate is generally limited by the fastest changing element in the gradient vector.

For the following investigation, we \secondrevision{adapt} the MPC parameterization \secondrevision{of $\tilde{E}_{A,\mathrm{ad}}$ by introducing the scaling factor $\xi$
\begin{align}
    \tilde{E}_{A,\mathrm{ad}} = (1 + \xi \, \theta_1) \cdot 0.95 \cdot E_{A,\mathrm{ad}}
\end{align}
with $\xi \in \{ 0.1, 1, 10\}$ scaling the gradient and Hessian of the expected cumulative reward with respect to $\theta_1$.
}
Choosing $\xi = 0.1$ represents the scenario of a well scaled optimization problem, whereas $\xi = 10$ represents a poorly scaled one.
In order to assess the robustness of the proposed Gauss-Newton approach with respect to this parameter curvature, we ran training runs for five random initial guesses, sampled uniformly according to \secondrevision{$\theta_{1,0} \sim \calU(-0.1\xi^{-1}, 0.1\xi^{-1})$} and \secondrevision{$\theta_{2,0} \sim \calU(-0.5, 0.5)$ for the low parameterization, and additionally $\theta_{3,0}, \ldots, \theta_{13,0} \sim \calU(-0.1, 0.1)$ for the medium parameterized policy. Due to computational intractability of the approximate Newton method for the high parameterization, we omit the investigation for the highly parameterized policy.}
In order to guarantee comparability and fairness, all approaches \secondrevision{within each parameterization level} use the same initial guesses.
All approaches use the same hyperparameters as in Section~\ref{subsec:CSTR_Scalability}.

\begin{table}
    \centering
    \caption{\secondrevision{Median (med.) and average (avg.) f}inal performance (expected cumulative reward) of the agents, trained with \secondrevision{Adam, Approx. Newton (AN) and Gauss-Newton (GN)} for five different initial parameter guesses on the differently scaled \secondrevision{low and medium} policy parameterizations.} \label{tab:Parameter_scale_final_results}
    \setlength{\tabcolsep}{6pt}
    \begin{tabularx}{\linewidth}{l@{\hspace{3pt}}c|CC|CC|CC}\toprule
          &  & \multicolumn{2}{c|}{\secondrevision{Adam}} & \multicolumn{2}{c|}{\secondrevision{AN}} & \multicolumn{2}{c}{\secondrevision{GN}}  \\
          &  \secondrevision{$\xi$} & \secondrevision{med.} & \secondrevision{avg.} & \secondrevision{med.} & \secondrevision{avg.} & \secondrevision{med.} & \secondrevision{avg.} \\\midrule \midrule
        \multirow{3}{*}{\rotatebox{90}{\secondrevision{low}}} 
        & \secondrevision{$0.1$} & \secondrevision{$-66.6$} & \secondrevision{$-68.9$} & \secondrevision{$-66.7$} & \secondrevision{$-68.0$} & \secondrevision{$-66.7$} & \secondrevision{$-68.0$} \\
        & \secondrevision{$1$}   & \secondrevision{$-66.7$} & \secondrevision{$-66.7$} & \secondrevision{$-66.7$} & \secondrevision{$-66.8$} & \secondrevision{$-66.7$} & \secondrevision{$-66.7$} \\
        & \secondrevision{$10$}  & \secondrevision{$-99.5$} & \secondrevision{$-86.6$} & \secondrevision{$-66.7$} & \secondrevision{$-66.7$} & \secondrevision{$-66.7$} & \secondrevision{$-66.7$} \\
        \midrule
        \multirow{3}{*}{\rotatebox{90}{\secondrevision{medium}}} 
        & \secondrevision{$0.1$} & \secondrevision{$-67.2$} & \secondrevision{$-74.6$} & \secondrevision{$-67.0$} & \secondrevision{$-74.8$} & \secondrevision{$-67.0$} & \secondrevision{$-74.3$} \\
        & \secondrevision{$1$} & \secondrevision{$-67.0$} & \secondrevision{$-74.8$} & \secondrevision{$-67.1$} & \secondrevision{$-72.2$} & \secondrevision{$-67.1$} & \secondrevision{$-74.1$} \\
        & \secondrevision{$10$} & \secondrevision{$-71.6$} & \secondrevision{$-71.6$} & \secondrevision{$-67.0$} & \secondrevision{$-67.3$} & \secondrevision{$-67.0$} & \secondrevision{$-67.0$} \\ \bottomrule
    \end{tabularx}
\end{table}
\begin{figure*}
    \centering
    \includegraphics[width = \linewidth]{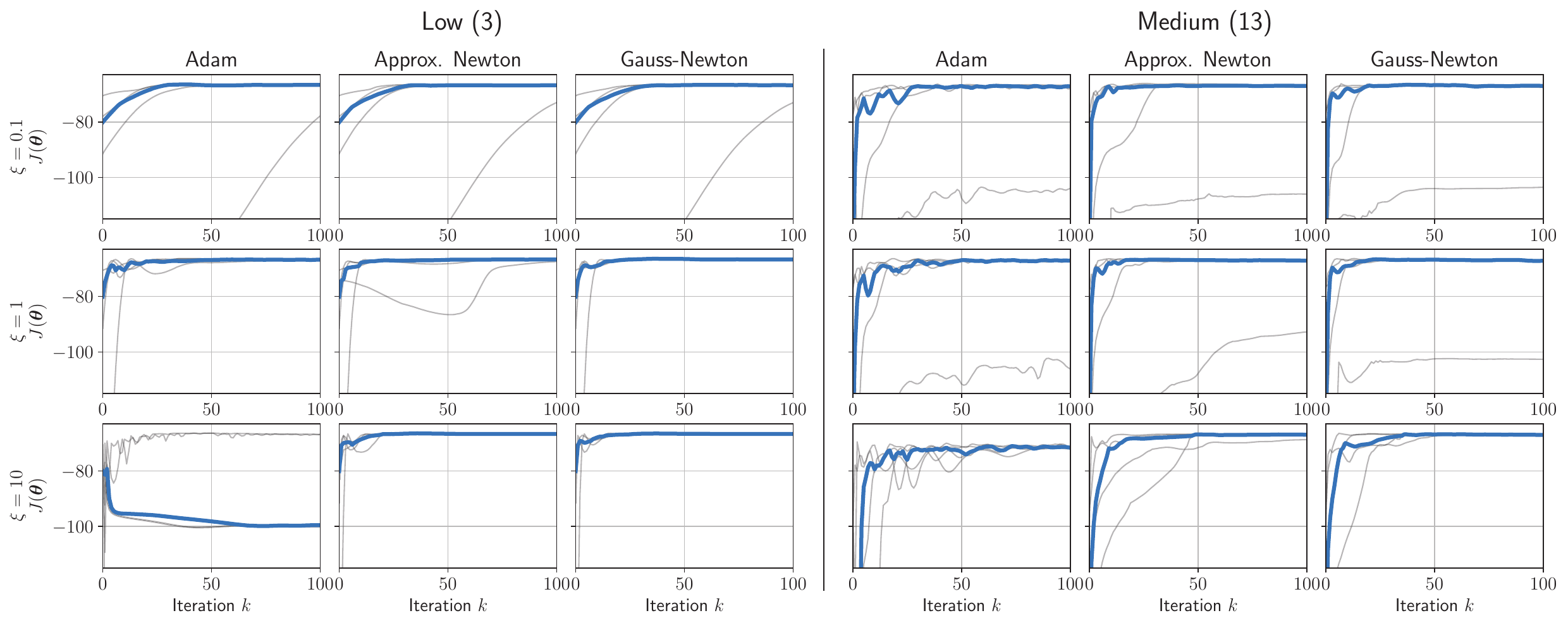}
    \caption{\secondrevision{Median expected cumulative reward (bold blue lines) and expected cumulative reward for five initial parameter guesses (gray lines) for the investigated policy optimization methods. Each row corresponds to a value of $\xi$. The left and right group of subplots represent the different parameterization levels (low and medium).}
    Adam uses $\alpha = 10^{-2}$. Both second order approaches use $\alpha = 10^{-1}$.
    } 
    \label{fig:ResilienceWRTParameterScale}
\end{figure*}
Figure~\ref{fig:ResilienceWRTParameterScale} shows the learning curves \secondrevision{for each initial parameter guess (gray lines) and the median performance (bold blue lines)} for three differently scaled \secondrevision{low and medium parameterized} agents.
\secondrevision{For the low parameterized policy, all} three algorithms perform well \secondrevision{$\xi=0.1$ and $\xi = 1$}.
\secondrevision{For $\xi = 10$} both second order approaches converge robustly towards the optimal policy for all initial parameter guesses, whereas the Adam optimizer fails to find the optimal policy in three out of five cases.
We found that choosing a smaller learning rate of $\alpha =10^{-3}$ remedied that behavior, however the learning progress was significantly slower compared to both second order approaches.
\secondrevision{The results for the medium parameterized policy, show a similar robust convergence for $\xi = 0.1$ and $\xi = 1$ for all but one initial guess. In early training stages, the median performance of both second order approaches improves more rapidly than Adam underlining the performance gains due to second order approaches. For $\xi =10$, both second order approaches converge robustly towards the optimal policy even for the challenging initial guess. On the other side, Adam also converges robustly, but to a different local optimum.}
The \secondrevision{median and average} final performance \secondrevision{of the differently trained} agents is summarized in Table~\ref{tab:Parameter_scale_final_results}.
\secondrevision{For all investigated scenarios the proposed Gauss-Newton approach achieves an average performance that is at least as good as Adam and even outperforms it most of the time, highlighting again the potential performance gains of second order approaches.}

We conclude that both second order approaches can be more resilient with respect to the sensitivity of the objective with respect to the parameter scale.
This can be of high interest especially when using MPC as policy in RL as the influence of the parameters on the closed loop performance can vary significantly depending on the parameter.

\end{reviewedblock}

\begin{reviewedblock}
\subsection{Comparison to deep reinforcement learning}
Finally, we also compare the performance of the trained MPC agents with a neural network~(NN) policy trained with the TD3 algorithm for deterministic policies~\citep{fujimotoAddressingFunctionApproximation2018}.
The implementation of the stable-baselines3~(SB3) toolbox is used~\citep{raffinStablebaselines3ReliableReinforcement2021}.
In order to guarantee a fair comparison, we performed a hyperparameter grid search.
All agents are trained for $10^6$ time steps, which is equivalent to $10^4$ episodes.
This is the same number of time steps as used for training of the agents in Section~\ref{subsec:CSTR_Scalability}.
The final choice of hyperparameters is given in Table~\ref{tab:Hyperparameters_TD3}.
\begin{table*}
    \centering
    \caption{Final hyperparameters for the TD3 algorithm. All other hyperparameters use the SB3 default implementation.}
    \label{tab:Hyperparameters_TD3}
    \begin{tabularx}{\linewidth}{CC|CC|CC} \toprule
         Hyperparameter & Value & Hyperparameter & Value & Hyperparameter & Value \\ \midrule \midrule
         Activation & gelu & Batch size & $1024$ & Buffer size & $ 10^5$\\
         Policy & $[64, 64]$ & Train frequency & $200$ episodes & Polyak coefficient & $10^{-1}$  \\
         Q-function & $[256, 128]$ & Gradient steps & $500$ & Policy delay & $10$    \\
         \bottomrule
    \end{tabularx}
\end{table*}

Figure~\ref{fig:SB3_comparison} shows the learning curve of the NN agent trained with the TD3 algorithm from SB3.
The learning curve is put into reference to the learning curves of the \secondrevision{low parameterized} MPC agents trained as described in Section~\ref{subsec:CSTR_Scalability}.
The figure shows that the MPC agents are already well initialized.
Still, the policy is fine-tuned during training, improving the performance even more.
The NN policy on the other side are not that well initialized.
Even after training, the final performance does not approach the performance of the initial MPC policies.
The performance of the different policies is summarized in Table~\ref{tab:PerformanceComparisonMPCvsNN}.
\begin{figure}
    \centering
    \includegraphics[width=\linewidth]{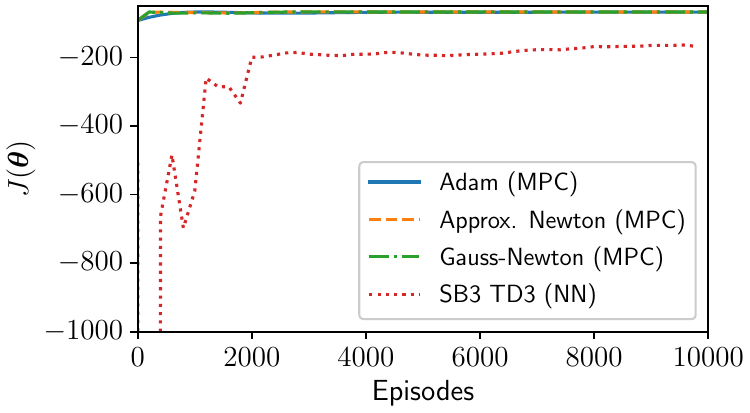}
    \caption{Expected cumulative reward over the observed training episodes for the considered MPC and NN agents.}
    \label{fig:SB3_comparison}
\end{figure}
\begin{table}
    \centering
    \caption{Expected cumulative reward after different numbers of episodes.}
    \label{tab:PerformanceComparisonMPCvsNN}
    \begin{tabularx}{\linewidth}{rCCCC}\toprule
         Episodes             &  $0$       & $5\,000$  & $10\,000$  \\ \midrule\midrule
         Adam                 & $-91.49$   & $-67.38$  & $-66.99$   \\
         Approx. Newton       & $-91.49$   & $-66.85$  & $-66.64$   \\
         Gauss-Newton         & $-91.49$   & $-66.86$  & $-66.66$   \\
         SB3 TD3              & $-506.41$  & $-192.80$ & $-167.90$   \\\bottomrule
    \end{tabularx}
\end{table}

To this end, we also exemplary compare the closed loop trajectories obtained via the untrained MPC agent, the MPC agent trained via the Gauss-Newton approach and the NN agent trained with the TD3 algorithm.
Figure~\ref{fig:Closed_loop} shows those closed loop trajectories.
The untrained MPC agent transitions rapidly from the initial condition towards the temperature setpoints.
However, it keeps a constant offset between the temperatures and the setpoints.
The upper constraint for $C_\mathrm{A}$ is slightly violated.
The trained MPC agent also transitions rapidly but, contrary to the untrained MPC agent, minimizes the offset between the temperatures and their setpoints.
Further, the constraint violation is decreased.
\secondrevision{Still they are not fully eliminated as constraint violations are only penalized via soft constraints. The trained MPC agent manages to find a policy that is a trade-off-between constraint violations, fast and accurate setpoint tracking, and low oscillations in the actions. The trained MPC agent is therefore optimal with respect to the provided reward. Note that these constraint violations can be reduced if the penalty in~\eqref{eq:CSTR_penalty_reward} is increased.}
Lastly, the NN policy transitions to the setpoints more conservatively than the MPC policies.
The setpoints are tracked well and the constraint is not violated due to the conservative behavior, \secondrevision{missing out on potential performance gains}.
In opposite to the MPC policies, the action trajectory is almost constant for $F_\mathrm{in}$ but \secondrevision{slightly} more oscillatory for $\dot{Q}$.
\begin{figure}
    \centering
    \includegraphics[width = \linewidth]{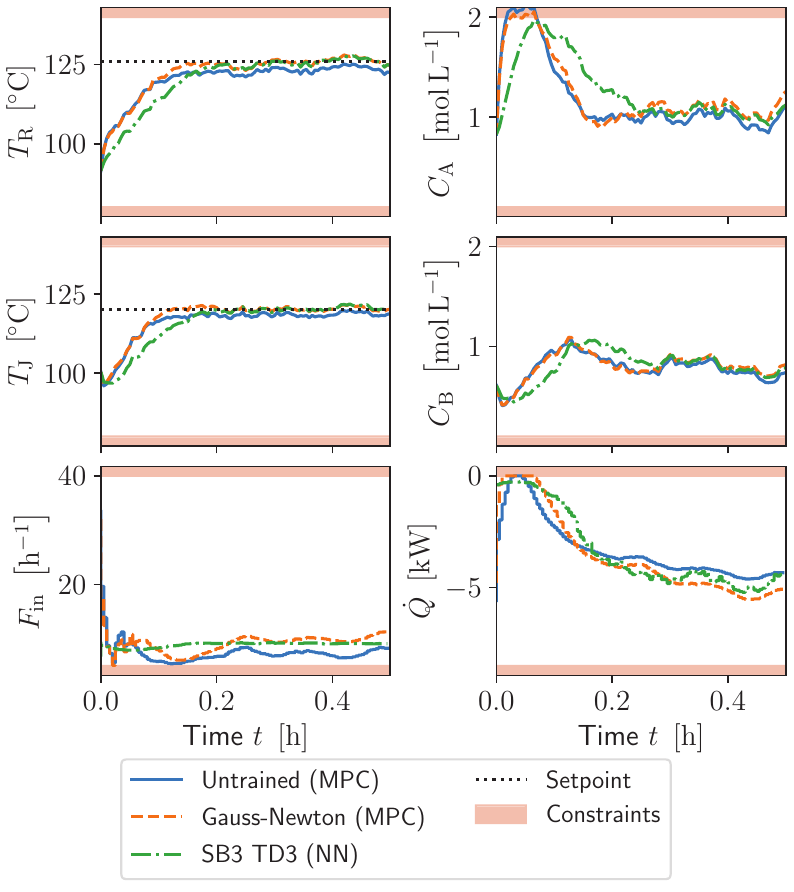}
    \caption{Closed-loop trajectories of the untrained MPC agent, the MPC agent trained via the proposed Gauss-Newton approach, and the TD3 agent.} \label{fig:Closed_loop}
\end{figure}
\end{reviewedblock}

\section{Conclusion and future work} \label{sec:Conclusion}
We presented a computationally efficient RL framework for MPC policies that enables computationally low-cost second-order policy optimization.
The key contribution is a Gauss-Newton approximation of the deterministic policy Hessian that avoids the computation of second-order policy derivatives while still enabling superlinear convergence.
The resulting approximation is well-suited for MPC policies as it can reduce the amount of expensive agent-environment interactions drastically due to its improved convergence behavior, while it simultaneously avoids to compute the computationally-expensive second-order NLP sensitivities.
Furthermore, we proposed a momentum-based exponentially moving averaging scheme for Hessian estimation that improves robustness against noise in local curvature samples.
\reviewed{We embed the Gauss-Newton Hessian approximation and the momentum-based averaging scheme in a trust-region constrained optimization problem}.
The proposed method improves stability during training and enables faster updates, which is beneficial for sample efficiency.

We rigorously proved the superlinear convergence behavior of the proposed Gauss-Newton approximation. This was confirmed empirically on an analytical example and a nonlinear CSTR benchmark.  %
Results show that the Gauss-Newton approximation can reduce the number of parameter updates compared to state-of-the-art first-order methods \reviewed{and state-of-the-art deep RL approaches.}
In addition, the proposed approach also reduces the total wall-clock time for training \reviewed{for increasingly more complex parameterized policies} compared to \reviewed{more elaborate} approximate policy Hessian methods while still maintaining robust learning behavior.
\reviewed{The proposed Gauss-Newton approach also shows resilience with respect to prescaling of the policy parameters. }

Future work will investigate scalability to higher-di\-men\-sio\-nal MPC policies, integration with robust MPC, and extensions to constrained Markov decision processes.

\printcredits

\bibliographystyle{cas-model2-names}

\bibliography{bibliography}

\end{document}

%% file: preamble.tex
\usepackage{blindtext}
\usepackage[normalem]{ulem}
\usepackage{tikz}
\usepackage{tabularx}
\newcolumntype{C}{>{\centering\arraybackslash}X}
\usepackage{pgfplots}
\usepackage{graphicx}
\usepackage{subcaption}
\pgfplotsset{compat=1.18}

\newtheorem{theorem}{Theorem}
\newtheorem{lemma}{Lemma}
\newtheorem{assumption}{Assumption}
\newtheorem{corollary}{Corollary}
\newdefinition{rmk}{Remark}
\newproof{pf}{Proof}
\newproof{pot}{Proof of Theorem \ref{thm}}

\newcommand\va{\boldsymbol{a}}

\newcommand\vg{\boldsymbol{g}}

\newcommand\vm{\boldsymbol{m}}

\newcommand\vp{\boldsymbol{p}}

\newcommand\vs{\boldsymbol{s}}

\newcommand\vu{\boldsymbol{u}}
\newcommand\vv{\boldsymbol{v}}
\newcommand\vw{\boldsymbol{w}}
\newcommand\vx{\boldsymbol{x}}

\newcommand\vz{\boldsymbol{z}}

\newcommand\mB{\boldsymbol{B}}

\newcommand\mD{\boldsymbol{D}}

\newcommand\mH{\boldsymbol{H}}

\newcommand\mM{\boldsymbol{M}}

\newcommand\mP{\boldsymbol{P}}

\newcommand\tT{\boldsymbol{\mathrm{T}}}

\newcommand\vzeta{\boldsymbol{\zeta}}

\newcommand\vtheta{\boldsymbol{\theta}}

\newcommand\vlambda{\boldsymbol{\lambda}}

\newcommand\vnu{\boldsymbol{\nu}}

\newcommand\vsigma{\boldsymbol{\sigma}}

\newcommand\vchi{\boldsymbol{\chi}}

\newcommand\mSigma{\boldsymbol{\Sigma}}

\newcommand\sE{\mathbb{E}}
\newcommand\sN{\mathbb{N}}
\newcommand\sR{\mathbb{R}}

\newcommand\calA{\mathcal{A}}

\newcommand\calL{\mathcal{L}}

\newcommand\calN{\mathcal{N}}

\newcommand\calS{\mathcal{S}}

\newcommand\calU{\mathcal{U}}

\newcommand\sequ{\boldsymbol{\mathrm{u}}}

\newif\ifshowrevisions
\showrevisionsfalse

\newcommand{\reviewed}[1]{%
  \ifshowrevisions
    \textcolor{blue}{#1}%
  \else
    #1%
  \fi
}
\newenvironment{reviewedblock}
{%
  \ifshowrevisions
    \color{blue}%
  \fi
}
{}

\newif\ifshowsecondrevisions
\showsecondrevisionsfalse

\newcommand{\secondrevision}[1]{%
  \ifshowsecondrevisions
    \textcolor{blue}{#1}%
  \else
    #1%
  \fi
}
\newenvironment{secondrevisionblock}
{%
  \ifshowsecondrevisions
    \color{blue}%
  \fi
}
{}